\documentclass[12pt]{article}

\usepackage[font=footnotesize]{caption}
\usepackage{longtable}
\usepackage{array}
\usepackage{enumerate}
\usepackage{float}
\usepackage{subfig}
\usepackage{multicol}
\usepackage{multirow}
\usepackage{epsfig}
\usepackage{graphicx}
\usepackage{pict2e}
\usepackage{color}
\usepackage{authblk}
\usepackage{cite}

\usepackage{amsmath,amsfonts,amssymb}
\usepackage{bbm}

\usepackage[utf8]{inputenc}
\usepackage[T1]{fontenc}
\usepackage[english]{babel}
\usepackage{xspace}
\usepackage{pifont}
\usepackage{collref}

\usepackage[pdftitle={Exact MSSM Models within F-theory},
    pdfauthor={Sven Krippendorf, Dami\'an Kaloni Mayorga Pe\~na, Paul-Konstantin Oehlmann, Fabian Ruehle},
    pdfsubject={},
    bookmarksopen, bookmarksnumbered, bookmarksopenlevel=2, colorlinks=false, linkcolor=blue, citecolor=blue, urlcolor=blue]{hyperref}

\newcommand{\U}[1]{\text{U(#1)}\xspace}
\newcommand{\SU}[1]{\text{SU(#1)}\xspace}
\newcommand{\SO}[1]{\text{SO(#1)}\xspace}
\newcommand{\E}[1]{\ensuremath{\text{E}_{#1}}\xspace}

\newcommand{\five}[0]{\ensuremath{\mathbf{5} }\xspace}
\newcommand{\ten}[0]{\ensuremath{\mathbf{10} }\xspace}
\newcommand{\tenb}[0]{\ensuremath{\overline{\mathbf{10}} }\xspace}

\newcommand{\fiveb}[0]{\ensuremath{\overline{\mathbf{5}} }\xspace}

\begin{document}


\thispagestyle{empty}
\begin{flushright} DESY-13-259\end{flushright}
\vskip 1 cm
\begin{center}
{\Large {\bf Rational F-theory GUTs without exotics}
}
\\[0pt]
\bigskip
\bigskip {
{\bf Sven Krippendorf$^{\,a,}$}\footnote{
E-mail: krippendorf@th.physik.uni-bonn.de},
{\bf Dami\'an Kaloni Mayorga Pe\~na$^{\,a,}$}\footnote{
E-mail: damian@th.physik.uni-bonn.de},
{\bf Paul-Konstantin~Oehlmann$^{\,a,}$}\footnote{
E-mail: oehlmann@th.physik.uni-bonn.de},
{\bf Fabian~Ruehle$^{\,b,}$}\footnote{
E-mail: fabian.ruehle@desy.de}
\bigskip }\\[0pt]
\vspace{0.23cm}
${}^{a}${\it Bethe~Center~for~Theoretical~Physics, Physikalisches~Institut~der~Universit\"at~Bonn,\\ Nussallee~12,~53115~Bonn,~Germany\\[12pt]
${}^{b}$ Deutsches Elektronen-Synchrotron DESY,\\ Notkestrasse~85,~22607~Hamburg,~Germany}

\bigskip
\end{center}

\begin{abstract}
\noindent We construct F-theory GUT models without exotic matter, leading to the MSSM matter spectrum with potential singlet extensions. The interplay of engineering explicit geometric setups, absence of four-dimensional anomalies, and realistic phenomenology of the couplings places severe constraints on the allowed local models in a given geometry. In constructions based on the spectral cover we find no model satisfying all these requirements. We then provide a survey of models with additional U(1) symmetries arising from rational sections of the elliptic fibration in toric constructions and obtain phenomenologically appealing models based on SU(5) tops.  
Furthermore we perform a bottom-up exploration beyond the toric section constructions discussed in the literature so far and identify benchmark models passing all our criteria, which can serve as a guideline for future geometric engineering.
\end{abstract}
\clearpage
\setcounter{page}{1}
\setcounter{footnote}{0}
\setcounter{tocdepth}{2}
\tableofcontents
\section{Introduction and summary of results}
\label{Sec:Introduction}
F-theory \cite{Vafa:1996xn} is a very fruitful patch in the string landscape to connect string theory to particle physics phenomenology. It has the potential to realize exceptional gauge groups and as such it provides a promising starting point for engineering grand unified theories (GUTs)\cite{Donagi:2008ca,Beasley:2008dc,Hayashi:2008ba,Beasley:2008kw}. Being the non-perturbative formulation of type~IIB string theory, F-theory is expected to inherit its successful mechanisms for moduli stabilization. Hence it allows for a realization of a bottom-up approach to string model building~\cite{Aldazabal:2000sa}.

Here we focus on engineering GUT theories in standard F-theory ${\cal N}=1$ compactifications on elliptically fibered Calabi--Yau fourfolds $Y_4$ with base space $B_3.$ The fiber degenerations over the base manifold $B_3$ encode the particle content as well as the corresponding interactions of the theory: If the fiber becomes singular at complex codimension one in the base, this singularity corresponds to gauge degrees of freedom for a non-Abelian symmetry of the ADE-type. Matter fields arise along so-called matter curves which are codimension two loci in the base. Finally, from codimension three singularities, i.e.~points at which the matter curves intersect, one can obtain the corresponding Yukawa couplings. 

We follow the usual model building strategy by considering a local four-cycle $S$ in the base, along which the fiber degeneration leads to an \SU{5} symmetry.\footnote{For model building approaches based on other grand unified groups see e.g.~\cite{Chen:2010ts,Antoniadis:2012yk,Tatar:2012tm,Callaghan:2012rv}.} Then one identifies the matter curves giving rise to $\five$- and $\ten$-plets and turns on fluxes along these curves to obtain a net chirality of the matter fields. Flux along the hypercharge direction is used to break the \SU5 to the Standard Model gauge group and to project out exotic triplet matter~\cite{Blumenhagen:2008zz,Beasley:2008kw,Donagi:2008kj}. We assume here that these local constructions can be embedded in a global compactification such that the hypercharge remains massless, as has been shown in explicit global realizations~\cite{Marsano:2009ym,Blumenhagen:2009yv,Grimm:2009yu,Marsano:2009wr,Chen:2010ts}. We also assume that the flux is such that problems with gauge coupling unification can be avoided~\cite{Blumenhagen:2008aw,Donagi:2008kj,Dolan:2011aq,Dolan:2011iu}.

As in many other model building approaches, additional \U1 symmetries are used in \hbox{F-theory} to generate a phenomenologically viable structure of operators~\cite{Dudas:2009hu,Marsano:2009wr,Marsano:2009gv,Dudas:2010zb,King:2010mq,Callaghan:2011jj,Dolan:2011iu}. In this setup, already the consistency conditions from four-dimensional anomaly cancellation place severe constraints on the local GUT construction~\cite{Marsano:2010sq,Palti:2012dd}. Furthermore, previous explorations (see for instance~\cite{Dudas:2009hu,Dolan:2011iu}) have shown that dangerous operators (e.g.~proton decay operators) tend to carry the same \U1 charges as desired couplings (e.g.~Yukawa couplings). As of now there is no consistent local setup with just the MSSM matter content and phenomenologically acceptable couplings. 

In the spectral cover approach to F-theory model building, the \U1 symmetries are realized as Cartan elements of an enhanced gauge group which can be maximally an $\E8$ symmetry. This gauge group is constraining the \U1 charges that can appear upon breakdown to an $\SU{5} \times \U1^N$ gauge group. More generally, \U1 symmetries in F-theory are associated with extra sections of the elliptic fibration \cite{Morrison:1996na,Morrison:1996pp}, which appear in addition to the generic zero section which specifies the base space. Significant progress has been made in constructing these sections and computing the \U1 charges for the matter representations and the singlets~\cite{Mayrhofer:2012zy,Braun:2013yti,Borchmann:2013jwa,Cvetic:2013nia,Braun:2013nqa,Cvetic:2013uta,Borchmann:2013hta,Cvetic:2013jta,Cvetic:2013qsa}. 
These new \U1 symmetries allow for a wider pattern of charges as compared to the \U1 symmetries arising in spectral cover constructions.

In this paper we explore the pattern of \U1 charges that can arise from the constructions outlined above and address the question of whether an exact MSSM construction can be obtained consistently within this class of F-theory compactifications. We systematically search for models that have the exact MSSM spectrum (up to singlet extensions), satisfy the four-dimensional anomaly cancellation conditions, and allow for phenomenologically viable couplings. In contrast to previous F-theory GUT model building, where one requires the MSSM fields to arise from complete \SU5 representations, we allow for incomplete multiplets -- a feature which commonly appears in the heterotic mini-landscape~\cite{Lebedev:2006kn}.

Following this strategy we revisit F-theory model building based on spectral cover constructions and take steps towards model building based on models with rational sections. We assume that the breakdown of the grand unified group in rational section models can proceed via hypercharge flux breaking, as in models based on the spectral cover.  We focus on constructions with up to two additional \U1 factors.\footnote{In the context of spectral covers it has been argued that the presence of more than two \U1 symmetries can spoil the flatness of the fibration and hence would lead to infinitely many undesired massless states.} For models with rational sections, we consider those with toric sections where the matter curves and the \U1 charges have been worked out. In this class we find examples that satisfy all constraints derived from imposing phenomenologically viable couplings and four-dimensional anomaly cancellation, apart from the $\U1^3$-anomaly discussed in~\cite{Palti:2012dd, Mayrhofer:2013ara}, whose presence depends on the embedding of the local model in the global compactification. Under the assumption that this anomaly can be canceled by an appropriate Green--Schwarz (GS) mechanism, we present one benchmark model in detail where the standard $\mathbbm{Z}_2$ matter parity can be realised. We then search for extensions of the currently available \U1 symmetries which follow a similar charge pattern as that appearing in rational section models, and ask which ones allow for a model satisfying all anomaly conditions, including the $\U1^3$ anomaly, and thus do not rely on an explicit realization of the GS mechanism mentioned above. In this way we find lamppost models whose geometric realization would be very intriguing.

The rest of this paper is organized as follows: In Section~\ref{Sec:Review} we briefly review the basics of model building in F-theory, with special focus on the spectral cover and the recent constructions with rational sections, which can be skipped by the experienced or only phenomenologically interested readers. In Section~\ref{sec:ModelBuilding} we explain our systematic search and present our results. We conclude in Section~\ref{sec:Conclusions} and provide a second benchmark model in Appendix~\ref{sec:AppendixBenchmarkModel2}.

\section{Review of F-theory model building techniques}
\label{Sec:Review}
F-theory can be thought of as a non-perturbative formulation of type IIB string theory (see \cite{Ibanez:2012zz,Weigand:2010wm,Heckman:2010bq,Maharana:2012tu} for recent reviews). It is a twelve-dimensional theory in which two dimensions are spanned by an elliptic curve. These two extra dimensions are not physical but can be regarded as a bookkeeping device that encodes in its complex structure the variation of the type IIB axio-dilaton
\begin{align*}
\tau=C_{0}+{\rm i} e^{-\phi}\, .
\end{align*}
We focus on elliptically fibered compactification spaces which can be described algebraically in terms of the short Weierstrass form
\begin{align}
\label{eq:WeierstrassShort}
y^2 = x^3 + fxz^4 + gz^6 \, ,
\end{align}
where $x,y,z$ are homogeneous coordinates in $\mathbbm{P}_{2,3,1}$, and $f$ and $g$ are sections of the base $B_3$. Thus the fibration over the base is completely determined by $f$ and $g$. The fiber degenerates at points where the discriminant
\begin{align}
\label{eq:Discriminant}
\Delta = 4 f^3 + 27 g^2
\end{align}
vanishes. At these points one can observe the usual monodromies for the axio-dilaton which occur at places where D7 branes sit. These fiber degenerations encode valuable information for particle physics. At complex codimension one in the base $B_3$, one obtains the gauge degrees of freedom: the corresponding gauge symmetry can be inferred from the vanishing order of $f$, $g$, and $\Delta$ according to the Kodaira classification \cite{Kodaira:1963}. A more intuitive picture on how the gauge symmetry arises is available from the duality to M-theory: One can smoothen out (blow-up) the singularities by replacing them with a tree of $\mathbbm{P}^1$'s. The intersection matrix of the $\mathbbm{P}^1$'s is the (negative of the) affine Cartan matrix of the ADE type Lie groups where the torus plays the role of the affine node. On the M-theory side, one can argue that in the limit where all $\mathbbm{P}^1$'s shrink to zero size, M2 branes wrapping the $\mathbbm{P}^1$'s become massless, giving then rise to the necessary degrees of freedom of an ADE type gauge theory. Similar arguments lead to the conclusion that at higher codimensions in the base one encounters the matter representations (codimension two) and their Yukawa interactions (at codimension three in the base).

\subsection{The spectral cover}
\label{sec:SpectralCoverReview}

For some F-theory compactifications, the spectral cover construction is a widely-used method to keep track of the various matter curves and gauge enhancements in explicit geometric backgrounds. In this approach, it is useful to consider the Tate form instead of the short Weierstrass form \eqref{eq:WeierstrassShort}
\begin{align}
\label{eq:TateForm}
y^2 = x^3 - a_1 xyz + a_2 x^2 z^2 - a_3 y z^3 + a_4 x z^4 + a_6 z^6\,,
\end{align}
where the $a_i$'s are again sections of the base.\footnote{The Tate form can be brought to the Weierstrass form by completing the square in $y$ and the cube in $x$ and subsequently redefining the fields.}
In this form,  $f$, $g$ and the discriminant $\Delta$ of \eqref{eq:Discriminant} are given in terms of the $a_i$ by
\begin{align}
\label{eq:KodairaTateMatch}
\begin{split}
f & = -\frac{1}{48}(\beta^2_2-24\beta_4 )\, , \qquad g=-\frac{1}{864}(-\beta^3_2+ 36 \beta_2 \beta_4 -216 \beta_6 )\, ,\\
\Delta & = \frac{\beta_2^2}{4}(\beta_4^2-\beta_2\beta_6)- 8 \beta_4^3 -27\beta_6^2 +9 \beta_2 \beta_4 \beta_6\,\,,
\end{split}
\end{align}
where
\begin{align}
\begin{split}
\beta_2 & = a_1^2 + 4a_2\, , \qquad \beta_4 = a_1 a_3 + 2a_4\, , \qquad \beta_6 = a_3^2 + 4a_6 \,.
\end{split}
\end{align}
The gauge symmetry one obtains for a given singularity can be inferred from the vanishing order of the $a_i$ according to the Tate classification \cite{Bershadsky:1996nh,Katz:2011qp}.
Since our focus is on \SU5 GUTs which are further broken down to the SM gauge group,
we are interested in a Tate or Weierstrass model that has an \SU5 singularity over a certain divisor $S$ (which is of complex codimension one in the base $B_3$)
\begin{align}
S : w = 0 \, .
\end{align}
To obtain the vanishing orders of an \SU5 singularity, we express the $a_i$'s in terms of $w$ and polynomials $b_i$ which do not contain overall factors of $w$ as
\begin{align}
\label{eq:TateGUTDivisor}
a_1 = b_5\, , \quad a_2 = b_4 w \, , \quad a_3 = b_3 w^2\, , \quad a_4 = b_2 w^3 \, , \quad a_6 = b_0 w^5 \,.
\end{align}
The $b_i$ are sections of the bundle $\eta-i c_1(S)$ where $\eta=6c_1(S)-t(S)$, with $c_1(S)$ and $-t(S)$ the first Chern class of the tangent and the normal bundle of $S$, respectively.

Inserting the parametrization \eqref{eq:TateGUTDivisor} into \eqref{eq:KodairaTateMatch}, we find for the discriminant $\Delta$
\begin{align}
\label{eq:DiscriminantGUTEnhancement}
\Delta = -w^5 \left[P_{10}^4 P_5 + w P_{10}^2(8b_4 P_5 + P_{10} R) + O(w^2)\right]
\end{align}
with
\begin{align}
\label{eq:PolynomialMatterCurves}
 P_5 &= (b_3^2 b_4 - b_2 b_3 b_5 + b_0 b_5^2) \,,\qquad P_{10} = b_5\,, \qquad R = -b_3^3 - b_2^2 b_5 + 4 b_0 b_4 b_5 \,.
\end{align}
The vanishing order of $w$ in $\Delta$ is increased to 6 or 7 on the subloci where $P_5$ or $P_{10}$ vanish, respectively. This hints at an \SU6 and \SO{10} gauge group enhancement.\footnote{Although the Kodaira classification is only valid in codimension one \cite{Esole:2011sm}, higher codimension singularities behave according to the naive expectations gained from the gauge group enhancements discussed here \cite{Marsano:2011hv}.} Matter fields arise at the intersections of $P_{5}=0$ ($P_{10}=0$) with $w=0$. The type of matter expected for each of these cases can be deduced from the decomposition of the adjoint representations into irreducible representations of $\SU5\times$\U1:
\begin{align}
P_{5}:\quad\mathbf{35}  & \supset  \mathbf{5} + \overline{\mathbf{5}}\,,\\
P_{10}:\quad\mathbf{45} & \supset \mathbf{10} + \overline{\mathbf{\mathbf{10}}}\,.
\end{align}
From the \SU6 enhancements one obtains the $\mathbf{5}$-plets whereas the $\mathbf{10}$-plets of \SU5 originate from the \SO{10} enhancements.

The spectral cover construction keeps the information of the various matter curves and gauge enhancements only in the vicinity of the GUT divisor $w=0$, which means that in \eqref{eq:TateGUTDivisor} one neglects possible terms involving $w$ in the $b_i$'s. To access this information one constructs a projective auxiliary (non-CY) threefold which contains the information about the discriminant locus in a hypersurface of this auxiliary space. In particular, we are interested in the following hypersurface with affine parameter $s$
\begin{align}
\label{eq:SpectralSurface}
 b_0 s^5+b_2s^3+b_3s^2+b_4s+b_5=0\, .
\end{align}
The correspondence with the matter curves is now as follows: from \eqref{eq:SpectralSurface} we see that e.g.\  the spectral surface at $s=0$ has $b_5=0$, i.e.~to the loci of \SO{10} enhancements where $P_{10}=0$  (cf.\ \eqref{eq:DiscriminantGUTEnhancement}, \eqref{eq:PolynomialMatterCurves}).

More generally, there is a local enhancement to \E8 and one then parameterizes the enhancements of \SU5 to \SO{10} or \SU6 in terms of a breaking of the fully enhanced \E8 into these groups. In order to obtain an \SU5 GUT surface in this construction, the structure group of the bundle that is responsible for the breaking of \E8 has to be \SU5. In order to distinguish the two \SU5 factors we call the latter $\SU5_\bot$. In terms of $\SU5\times\SU5_\bot$, the adjoint decomposes as
\begin{align}
\label{eq:AdjointDecomposition}
\mathbf{248}\; \rightarrow\; \mathbf{(24,1_\bot)} + \mathbf{(1,24_\bot)} + \mathbf{ (\overline{5}, 10_\bot)}+ \mathbf{(5,\overline{10}_\bot)}+   \mathbf{(10,5_\bot)}+ \mathbf{(\overline{10},\overline{5}_\bot)}\,.
\end{align}
The \SU5 matter can then be identified via their charges under the $\U1^4\subset\SU5_\bot$ Cartan subalgebra, which we characterize by the 5 weights $t_i$, $i=1,\ldots,5$. The $t_i$'s fulfill the tracelessness condition $\sum_i t_i = 0$ to ensure that the structure group is \SU5 rather than \U5. From~\eqref{eq:AdjointDecomposition} we see that the $\ten$ matter of \SU5 is paired with the $\five_\bot$ of $\SU5_\bot$, the $\overline{\five}$ GUT matter is paired with the $\ten_\bot$, and the GUT singlets are paired with the adjoint of $\SU5_\bot$. Thus the corresponding five $\ten$-curves $\Sigma_{\ten_i}$, ten $\five$-curves $\Sigma_{\five_{ij}}$, and 24 singlet curves $\Sigma_{\mathbf{1}_{ij}}$ are given in terms of the $t_i$ via

\label{eq:SCrepres}
 \begin{align}
\begin{split}  \Sigma_{\ten_i} &:\qquad t_i=0\,, \\
  \Sigma_{\fiveb_{ij}}&:\qquad (t_i + t_j)=0\,,\quad i\neq j \\
  \Sigma_{\mathbf{1}_{ij}}&:\qquad\pm(t_i - t_j)\,.
\end{split} \end{align}
The $b_i$'s can be expressed as symmetric polynomials of degree $i$ in the $t_i$'s. In particular, $b_5=P_{10}= t_1 t_2 t_3 t_4 t_5$. Thus, on the $\Sigma_{\ten_i}$-curves, one of the $t_i$ vanishes, leaving one Cartan generator of $\SU5_\bot$ unbroken. This reduces the structure group and allows for an enhancement from \SU5 to \SO{10}. Likewise, by plugging the $t_i$'s into $P_5$, we find at the loci exhibiting $SU(6)$ enhancement that $(t_i + t_j) = 0$, $i\neq j$. Furthermore, we have $b_1=t_1+t_2+t_3+t_4+t_5=0$. This is in accordance with the fact that there is no $a_5$-term in the Tate form \eqref{eq:TateForm}, which would have a coefficient $b_1$ in \eqref{eq:TateGUTDivisor}.

Albeit concise, the spectral cover loses the information about possible monodromies for the $t_i$'s. In other words, some of the matter curves might be identified identified away from the $\E8$ point, leading to fewer curves. Thus, depending on the monodromies, there can be zero to four extra \U1 symmetries appearing. Each \U1 is related to a polynomial of smaller degree $n_j$ in the affine parameter $s$, which can be factored out of the spectral cover equation \eqref{eq:SpectralSurface}, such that all $n_j$'s sum to five and that the term $s^4$ does not occur. For one \U1, the spectral cover has to split into two polynomials, which leaves us with the two possibilities of either a linear and a quartic polynomial ($4+1$ factorization) or a quadratic and a cubic polynomial ($3+2$ factorization). Since there is only one \U1, the $t_i$'s are identified by monodromies (e.g.\ $t_1\leftrightarrow t_2\leftrightarrow t_3$ and $t_4\leftrightarrow t_5$ in the $3+2$ factorization). For the case of two \U1 symmetries one proceeds in a similar way, finding a $2+2+1$ factorization and a $3+1+1$ factorization. We summarize the matter fields and their \U1 charges, which can be calculated from the possible \E8 embeddings, in Table~\ref{tab:Splittings}.

\begin{table}[t]
\centering
\begin{tabular}[b]{cccc}
\subfloat[4+1 factorization \label{tab:T1}]{
\renewcommand{\arraystretch}{1}
\begin{tabular}[b]{|c|c|}
\hline
 Curve & $q$ \\
\hline
$\mathbf{10}_{1}$ & $1$ \\
$\mathbf{10}_{5}$ & $-4$ \\
$\overline{\mathbf{5}}_{11}$ & $2$ \\
$\overline{\mathbf{5}}_{15}$ & $-3$ \\
\hline
\end{tabular}
~~~~
}
&
\subfloat[3+2 factorization \label{tab:T2}]{
\renewcommand{\arraystretch}{1}
\begin{tabular}[b]{|c|c|}
\hline
 Curve & $q$ \\
\hline
$\mathbf{10}_{1}$ & $2$ \\
$\mathbf{10}_{4}$ & $-3$ \\
$\overline{\mathbf{5}}_{11}$ & $4$ \\
$\overline{\mathbf{5}}_{14}$ & $-1$ \\
$\overline{\mathbf{5}}_{44}$ & $-6$ \\
\hline
\end{tabular}
~~~~~
}
&
\subfloat[2+2+1 factorization \label{tab:T3}]{
\renewcommand{\arraystretch}{1}
\begin{tabular}[b]{|c|c|c|}
\hline
 Curve & $q_1$ & $q_2$ \\
\hline
$\mathbf{10}_{1}$ & $1$ & $5$\\
$\mathbf{10}_{3}$ & $1$ & $-5$\\
$\mathbf{10}_{5}$ & $-4$ & $0$\\
$\overline{\mathbf{5}}_{11}$ & $2$ & $10$ \\
$\overline{\mathbf{5}}_{13}$ & $2$ & $0$ \\
$\overline{\mathbf{5}}_{33}$ & $2$ & $-10$\\
$\overline{\mathbf{5}}_{15}$ & $-3$ & $5$ \\
$\overline{\mathbf{5}}_{35}$ & $-3$ & $-5$\\
\hline
\end{tabular}
}
&
\subfloat[3+1+1 factorization \label{tab:T4}]{
\renewcommand{\arraystretch}{1}
\begin{tabular}[b]{|c|c|c|}
\hline
 Curve & $q_1$ & $q_2$ \\
\hline
$\mathbf{10}_{1}$ & $2$ & $0$\\
$\mathbf{10}_{4}$ & $-3$ & $5$\\
$\mathbf{10}_{5}$ & $-3$ & $-5$\\
$\overline{\mathbf{5}}_{11}$ & $4$ & $0$ \\
$\overline{\mathbf{5}}_{14}$ & $-1$ & $5$ \\
$\overline{\mathbf{5}}_{15}$ & $-1$ & $-5$\\
$\overline{\mathbf{5}}_{45}$ & $-6$ & $0$ \\
\hline
\end{tabular}
}
\end{tabular}
\caption{\U1 charges of $\SU5_{\text{GUT}}$ representations for different factorizations with up to two \U1 factors. The indices specify the $\SU5_\bot$ Cartan weights according to \eqref{eq:SCrepres}.}
\label{tab:Splittings}
\end{table}

\subsection{Compactifications with multiple sections}
\label{sec:RationalSectionsReview}

Motivated by obtaining more general \U1 charge assignments which are not constrained by the embedding into \E8 as in the spectral cover, we briefly review how \U1 symmetries can appear more generally in F-theory.

In contrast to non-Abelian gauge groups which arise from local degenerations of the elliptic fiber \cite{Vafa:1996xn},
 \U1 gauge factors are related to the presence of additional sections of the elliptic fibration \cite{Morrison:1996na,Morrison:1996pp}. In general they depend on global properties of the compactification space.

Compactification spaces which are used in F-theory come with a holomorphic section that specifies the base of the fibration\footnote{See \cite{Cvetic:2013nia} for a discussion with rational sections instead of holomorphic ones.},
which is known as the zero section. In terms of the Weierstrass model~\eqref{eq:WeierstrassShort}, it corresponds to the point $O=[x:y:z]=[\lambda^2:\lambda^3:0]$ on the torus. In \cite{Andreas:1999ty,Mayrhofer:2012zy}, additional sections were constructed by factorizing the Weierstrass equation. The novel property is that the corresponding \U1 factors do not necessarily descend from an \E8. These extra sections need not be holomorphic. They simply have to be rational, so that in principle they can wrap entire fiber components at base codimensions greater than one.

The set of sections is known to form the so-called Mordell--Weil group of the compactification~\cite{opac-b1103238}. The rank of this group matches the number of \U1 factors since sections are related to harmonic two-forms and the \U1 gauge fields are obtained from expansions of the M-theory three-form in a suitable basis of these.
In order to study the Mordell--Weil group one fixes a point $O$ on the torus (the one related to the zero section). Addition of two rational points $P$ and $Q$ on the torus is then defined by translating $P$ by the element of $\U1^2$ associated to $Q$. In terms of the elliptic curve $E$, the group operation is defined as follows: A line $\mathbbm{P}^1\subset\mathbbm{P}^2$ intersects the elliptic curve in three points (counted with multiplicities). In this way, given two points $P$ and $Q$, one can define a third point $R$ as the point of collision of the line $PQ$ with $E$. Group addition is now defined by identifying the point $P+Q$ with the point of collision of the line $OR$ with $E$. If we consider an elliptic curve to be specified as a cubic in $\mathbbm{P}^2$ with coefficients from the field $\mathbbm{Q}$, we see that the rational points on the elliptic curve indeed form a group, the so-called Mordell--Weil group.\footnote{See appendix A of \cite{Braun:2013nqa} for an example.} This can be applied to the previous discussion by mapping the cubic in $\mathbbm{P}^2$ to the Weierstrass form \eqref{eq:TateForm} in $\mathbbm{P}_{2,3,1}$. If one uses the function field $\mathbbm{C}[s]$ of meromorphic functions, the rational points in $\mathbbm{C}[s]$ are sections of the bundle of elliptic curves over the complex line with coordinate~$s$.

In general, the extra sections do not necessarily intersect the same irreducible fiber component as the zero section; instead, they can intersect any of the $\mathbbm{P}^1$'s glued into the fiber to smoothen out the singularities over the GUT divisor. Of course, any section can be chosen as the zero section (corresponding to a choice of $O$ on the torus). The $\mathbbm{P}^1$ that is intersected by the zero section fixes by definition the affine node $A_0$ of the extended Dynkin diagram. In the case of one additional section that intersects a different $\mathbbm{P}^1$ corresponding to one of the other four nodes $A_i$ (counted clockwise in our convention) of the Dynkin diagram, the results agree with a so-called $i-(5-i)$ split. If the extra section also intersects the affine node, this gives rise to a $5-0$ split. In the presence of two or more \U1 symmetries, one has to choose again a zero section and subsequently specify the node $A_i$ that is intersected by each additional section independently, i.e.\ one has to specify an $i-(5-i)$ split for each \U1 factor independently. The split is closely related to the charges of the matter fields under the corresponding \U1 symmetry, which are fixed by the Shioda map. Here we omit its construction which can be found for example in \cite{Morrison:2012ei}, and simply quote the results. The Shioda map contains the intersection numbers for the sections and the irreducible fiber components as part of its input. Furthermore, it involves the inverse of the Cartan matrix of \SU5, which implies that the \U1 charges will be quantized in multiples of $1/5$. It is common to choose a \U1 normalization such that all charges are integral. This is the reason why, under a given split, the charges of \textit{all} \ten-plets, \fiveb-plets and singlets are subjected to the following relations:
\begin{equation}
q_{\fiveb}=Q_{\fiveb}+5\mathbb{Z}\,,\quad q_{\ten}=Q_{\mathbf{10}}+5\mathbb{Z}\,,\quad q_{\mathbf{1}}=0+5\mathbb{Z}\, ,\label{eq:SplitCharge}
\end{equation}
where $Q_{\fiveb}$ and $Q_{\mathbf{10}}$ are given in Table~\ref{tab:Charges} for all possible splits. Note that in all cases the charges for the fields in the different spectral cover factorizations also obey the relations \eqref{eq:SplitCharge}. For example, the charges in the 4+1 factorization match those of a 3-2 splitting, cf.\ Tables~\ref{tab:Splittings}~and~\ref{tab:Charges}. One can also show that all \SU5 allowed operators carry \U1 charges divisible by five due to the structure of the Shioda map.
\begin{table}[t]
\centering
\renewcommand{\arraystretch}{1}
\begin{tabular}{|c|ccccc|}
 \hline
 Split & 5-0 & 4-1 & 3-2 & 2-3 & 1-4 \\
 \hline
  $Q_{\fiveb}$ & 0 & 1 & 2 & 3 & 4 \\
  $ Q_{\ten}$ & 0 & 3 & 1 & 4 & 2 \\
 \hline
\end{tabular}
\caption{Charge assignments for \fiveb- and \ten-curves for all possible splittings.}
\label{tab:Charges}
\end{table}
\subsection{Explicit examples from toric constructions}
\label{sec:ToricSectionsReview}

Sections that can be described by toric geometry constitute a tractable subset of all rational sections. We will focus on them in the following. There are 16 possible ways to write the elliptic fiber as a hypersurface in a toric ambient space \cite{Kreuzer:1995cd}, each of which is characterized by its toric diagram or polygon. The toric sections of these polygons give rise to (the toric part of) their corresponding Mordell--Weil group\footnote{In addition, there can be non-toric sections (i.e.\ sections that are not simply given by setting one fiber coordinate to zero) giving rise to further non-toric \U1 factors.} and have been analyzed in \cite{Braun:2013nqa}. These polygons can lead to up to three toric \U1 symmetries.

\begin{figure}[t]
  \centering
  \setlength{\unitlength}{0.1\textwidth}
  \begin{picture}(8,2.5)
    \put(1,0.2){\includegraphics[width=0.65\textwidth]{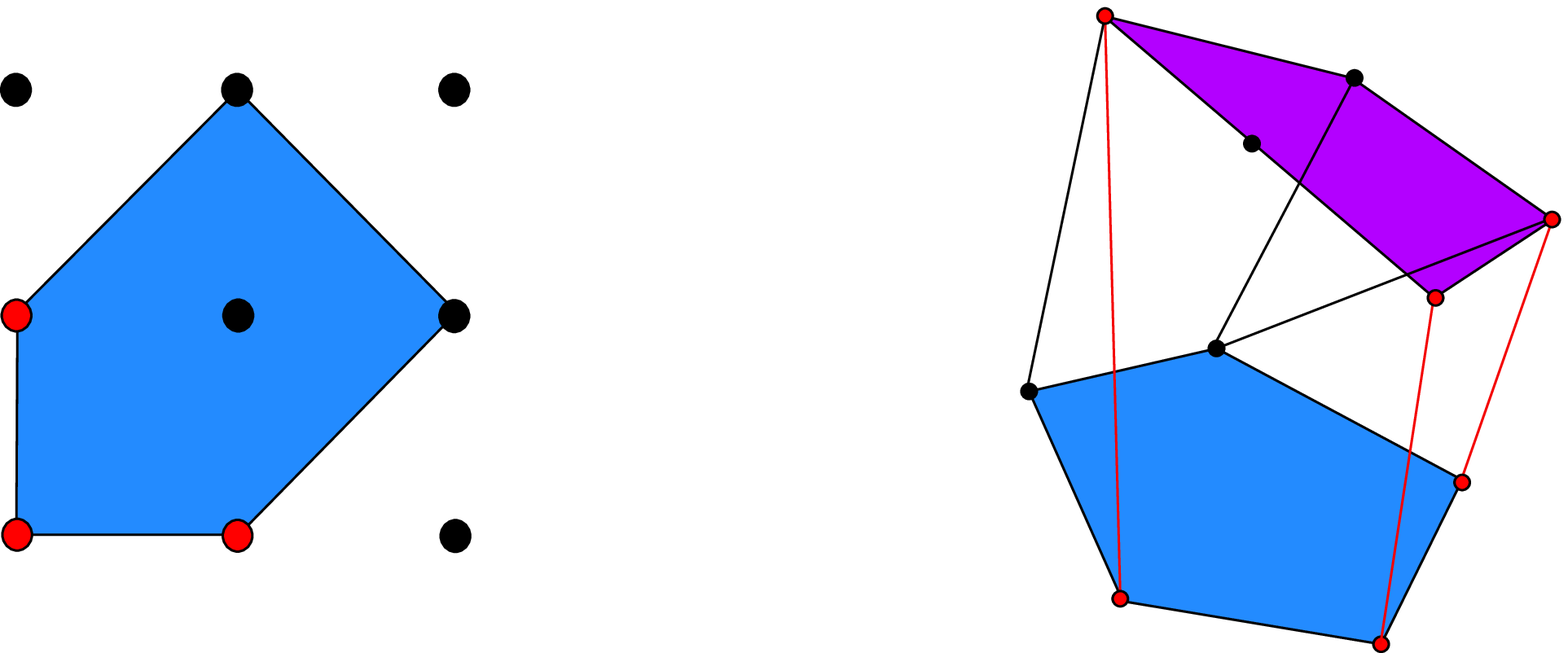}}
    \put(0.65,1.55){$s_1$}
    \put(0.75,0.55){$u$}
    \put(1.75,0.55){$s_0$}
     \put(1.75,2.60){$w$}
     \put(2.82,1.55){$v$}
     \put(4.5,2.8){$z=1$}
     \put(4.1,1.3){$z=0$}
  \end{picture}
  \vspace{0.5cm}
  \centerline{(a)\hspace{7cm}(b)}
  \vspace{0.5cm}
  \begin{picture}(8,2)
    \put(0,0.0){\includegraphics[width=0.8\textwidth]{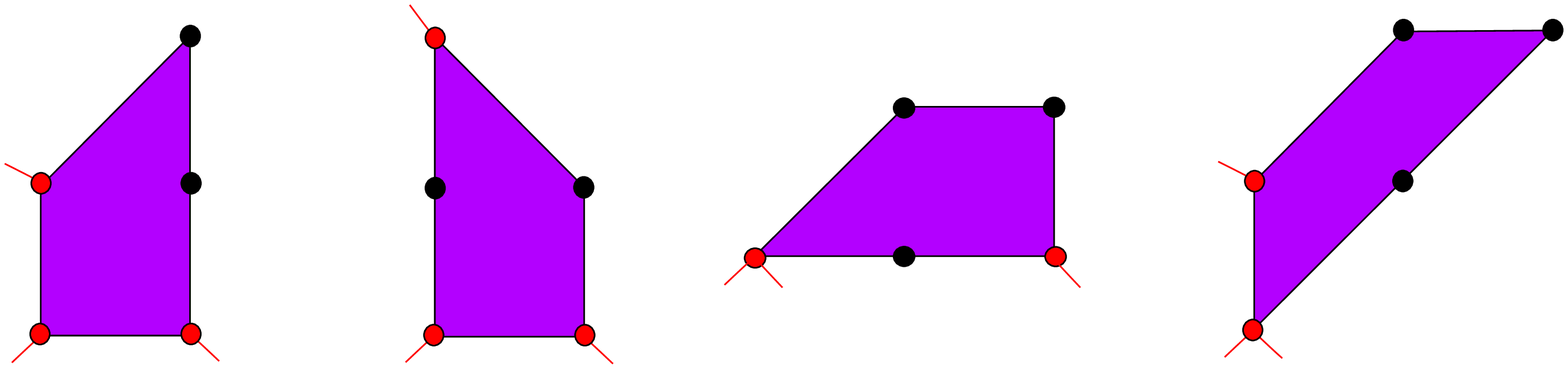}}
    \put(-0.30,1.05){$\sigma_1$}
    \put(-0.20,-0.15){$\sigma_2$}
    \put(1.10,-0.15){$\sigma_0$}
    \put(1.90,1.95){$\sigma_1$}
    \put(1.90,-0.15){$\sigma_2$}
    \put(3.10,-0.15){$\sigma_0$}
     \put(3.50,0.20){$\sigma_1$}
    \put(4.00,0.20){$\sigma_2$}
    \put(5.50,0.20){$\sigma_0$}
    \put(5.95,1.05){$\sigma_1$}
    \put(6.10,-0.15){$\sigma_2$}
    \put(6.55,-0.15){$\sigma_0$}
  \end{picture}
  \centerline{(c) $\tau_{5,1}$\hspace{1.8cm}(d) $\tau_{5,2}$\hspace{2.8cm}(e) $\tau_{5,3}$\hspace{3.0cm}(f) $\tau_{5,3}$}
  \caption{(a) Toric diagram for the polygon $F_5$. It has the sections $\sigma_0$: $s_0=0$ (the zero section), $\sigma_1$: $s_1=0$ and $\sigma_2$: $u=0$, which are marked as red points in the diagram. (b) The polygon is set as the basis for the \SU5 top at $z=0$. The intersections of the sections with the tree of $\mathbb{P}^1$'s at $z=1$ (see the red lines in the diagram) serve to compute the charges of the fields via the Shioda map. From the intersections one can also deduce the splitting for each of the inequivalent flat \SU5 tops allowed for $F_5$: (c) 2-3, 1-4 (d) 3-2, 1-4 (e) 2-3, 2-3 and (f) 1-4, 5-0.}
  \label{fig:F5Top}
\end{figure}

The polygon gives us a description of the elliptic fiber and its sections. Next we have to include the information about the possible ways for desingularizing the \SU5 degenerations. In order to do so, we combine two polygons to form a three-dimensional polyhedron known as a \emph{top}. We first place the fiber polygon into the plane $z=0$. Parallel to it, at $z=1$, we then introduce another polygon whose integer boundary points correspond to the five nodes of the affine Dynkin diagram of \SU5 (see Figure \ref{fig:F5Top}\,(b) for an example). In this way, the facet at $z=0$ encodes the generic fiber, whereas the one at $z=1$ encodes its \SU5 resolution. However, the top completion is not unique and for every of the 16 polytopes there can be multiple tops\footnote{Note that tops can be related by symmetries or might not lead to a flat fibration (which means that there will be an infinite tower of massless fields), thus reducing the amount of viable tops.} \cite{Braun:2013nqa}.

As already mentioned, in order to compute the \U1 charges for the fields, one needs to find the intersections of the toric sections with the irreducible fiber components. Using the top, these intersections can simply be read off from the edges that are shared between a vertex of the fiber polygon that corresponds to the toric section and a vertex of the other polygon that corresponds to an irreducible $\mathbbm{P}^1$.
To exemplify this, the reader is referred to Figure \ref{fig:F5Top}, where details of the top $\tau_{5,2}$ are given. The red lines correspond to the intersections of the sections with the irreducible fibers. The upper facet of the top in Figure \ref{fig:F5Top}\,(b) is shown in Figure \ref{fig:F5Top}\,(d), from which one can see that this intersection pattern is consistent with a splitting of the form 3-2 and 1-4 for the first and second \U1 symmetries. 
The singlet fields will emerge from codimension two loci where the fiber degenerates to two irreducible components (i.e. the extended Dynkin diagram of \SU2). Analogously as for the matter curves, the charges of the singlets are computed from the intersection pattern of the sections with these two fiber components \cite{Morrison:2012ei}.

A very important feature about these constructions is that the charges of the \ten-plets under toric \U1 symmetries are very constrained: The \ten matter curves are given in terms of the triangulation of the facet of the polygon at height one. We want to choose this polygon such that it does not contain an interior point in order to maintain flatness of the fibration. Due to this we are left, up to isomorphisms, with only one such polygon (see Figure \ref{fig:F5Top} (c)-(f)). This polygon admits two different triangulations corresponding to two possible degenerations of the $\SU5$ to an $\SO{10}$ Kodaira fiber. However, the choice is fixed by the top to be universal over the whole GUT divisor, such that the \U1 charges of all \ten matter curves coincide. This is in contrast to the \five matter curves where the different degenerations can occur over different codimension two loci. Note that more general situations where different \ten-curves carry different \U1 charges can be obtained by taking complete intersections instead of hypersurfaces. However, these constructions have not been studied in the literature up to now such that we do not include this possibility in our subsequent analysis.

In \cite{Borchmann:2013hta, Cvetic:2013uta} five of the 16 possible reflexive polytopes were completed using all inequivalent tops that lead to \SU5. Out of these five polygons only the polygon $F_5$, corresponding to $dP_2$, exhibits two U(1) gauge factors while the others exhibit only one or less toric U(1) symmetries. After the completion to \SU5, matter curves and their \U1 charges have been calculated for every of these tops. As $F_5$ will be of main phenomenological interest we show the matter content and U(1) charges for its four possible inequivalent tops that can lead to a flat fibration in Table~\ref{tab:TopCharges}. There we also include the singlet spectrum, which is universal for all of these tops, as it depends only on the base polygon $F_5$.

So far we have not specified the whole CY fourfold or the base space. Thus, as a next step we would have to complete the top to a polygon which describes the complete CY fourfold. Indeed, whether or not a given base polytope can be combined with any of the tops in such a way that the fibration is flat depends on the choice of the base \cite{Lawrie:2012gg,Braun:2013nqa,Borchmann:2013hta}. However, a detailed study of the possible base spaces that complete the top is beyond the scope of this paper, so that we will simply assume that there exists a choice for the base such that the fibration is flat with the maximum amount of matter curves.
Given this assumption, the other properties like the amount of toric \U1 symmetries and non-Abelian gauge factors or the \U1 charges can be studied from the top alone without the need of specifying a base.
Since it will be our main concern to satisfy the anomaly cancellation constraints, it is sufficient for us to know the number of \U1 symmetries, how many curves we can expect, and what their \U1 charge pattern can be. As we will show in the following, anomaly cancellation places very strong constraints on the fluxes.
\begin{table}[t]
\centering
\begin{tabular}[b]{cccc}
\subfloat[Top $\tau_{5,1}$. \label{tab:RS1}]{
\renewcommand{\arraystretch}{1}
\begin{tabular}[b]{|c|c|c|}
\hline
 Curve & $q_1$ & $q_2$ \\
\hline
$\mathbf{10}_{1}$ & $-1$ & $2$ \\
$\overline{\mathbf{5}}_1$ & $3$ & $-1$ \\
$\overline{\mathbf{5}}_{2}$ & $-2$ & $4$ \\
$\overline{\mathbf{5}}_{3}$ & $-2$ & $-6$ \\
$\overline{\mathbf{5}}_{4}$ & $3$ & $4$ \\
$\overline{\mathbf{5}}_{5}$ & $-2$ & $-1$ \\
\hline
\end{tabular}
}
&\hspace{-2mm}
\subfloat[Top $\tau_{5,2}$. \label{tab:RS2}]{
\renewcommand{\arraystretch}{1}
\begin{tabular}[b]{|c|c|c|}
\hline
 Curve & $q_1$ & $q_2$ \\
\hline
$\mathbf{10}_{1}$ & $1$ & $2$ \\
$\overline{\mathbf{5}}_1$ & $-3$ & $4$ \\
$\overline{\mathbf{5}}_{2}$ & $-3$ & $-6$ \\
$\overline{\mathbf{5}}_{3}$ & $-3$ & $-1$ \\
$\overline{\mathbf{5}}_{4}$ & $2$ & $4$ \\
$\overline{\mathbf{5}}_{5}$ & $2$ & $-1$ \\
\hline
\end{tabular}
}
&\hspace{-2mm}
\subfloat[Top $\tau_{5,3}$. \label{tab:RS3}]{
\renewcommand{\arraystretch}{1}
\begin{tabular}[b]{|c|c|c|}
\hline
 Curve & $q_1$ & $q_2$ \\
\hline
$\mathbf{10}_{1}$ & $-1$ & $-1$ \\
$\overline{\mathbf{5}}_1$ & $3$ & $-2$ \\
$\overline{\mathbf{5}}_{2}$ & $-2$ & $-7$ \\
$\overline{\mathbf{5}}_{3}$ & $-2$ & $3$ \\
$\overline{\mathbf{5}}_{4}$ & $3$ & $3$ \\
$\overline{\mathbf{5}}_{5}$ & $-2$ & $-2$ \\
\hline
\end{tabular}
}
&\hspace{-2mm}
\subfloat[Top $\tau_{5,4}$. \label{tab:RS4}]{
\renewcommand{\arraystretch}{1}
\begin{tabular}[b]{|c|c|c|}
\hline
 Curve & $q_1$ & $q_2$ \\
\hline
$\mathbf{10}_{1}$ & $2$ & $0$ \\
$\overline{\mathbf{5}}_1$ & $4$ & $5$\\
$\overline{\mathbf{5}}_{2}$ & $4$ & $0$ \\
$\overline{\mathbf{5}}_{3}$ & $-1$ & $5$ \\
$\overline{\mathbf{5}}_{4}$ & $-1$ & $-5$ \\
$\overline{\mathbf{5}}_{5}$ & $-1$ & $0$ \\
\hline
\end{tabular}}
\\
\multicolumn{4}{c}{
\subfloat[Singlet spectrum and charges. \label{tab:SS}]{
\renewcommand{\arraystretch}{1}
~~~~\begin{tabular}{|c|cccccc|}
\hline
Curve & $\mathbf{1}_{1}$ & $\mathbf{1}_{2}$ & $\mathbf{1}_{3}$ & $\mathbf{1}_{4}$ & $\mathbf{1}_{5}$ &
$\mathbf{1}_{6}$ \\ \hline
$q_1$ & $-5$ & $5$ & $5$ & $-5$ & $0$ & $0$ \\

$q_2$ & $5$ & $0$ & $10$ & $-5$ & $-10$ & $5$ \\
\hline
\end{tabular}~~~~}
}
\end{tabular}
\caption{\U1 charges of the four inequivalent tops based on the fiber polygon $F_5$. The singlet charges are the same for all tops.}
\label{tab:TopCharges}
\end{table}

\subsection{Fluxes and anomaly cancellation}
\label{Sec:DP}

Before switching on fluxes the matter curves support vector-like states. Flux is a crucial ingredient for model building since it is needed in order to obtain a chiral spectrum. However, one has to ensure the absence of anomalies in chiral spectra. This constrains the fluxes as we will review in the following.

One can distinguish between the following cases: in the first case, the flux leaves the \SU5 symmetry unbroken \cite{Hayashi:2008ba,Braun:2011zm,Krause:2011xj,Grimm:2011fx,Kuntzler:2012bu}. This can be used to get a net number of \five\ (\fiveb) or \ten\ (\tenb) of \SU5 living at the different matter curves.
The second possibility is to break the GUT symmetry by switching on flux along the GUT surface. In order to break \SU5 down to the SM gauge group, this flux is chosen to be proportional to the hypercharge \U1 generator (within the \SU5).\footnote{However, one has to ensure that the hypercharge generator does not get a St\"uckelberg mass, i.e.\ it must not couple to the closed RR sector. This is guaranteed by putting the hypercharge flux along a cycle which is trivial in the homology of the base but not in the homology of the GUT surface \cite{Buican:2006sn}.}

Even though the previous constraint depends strongly on the fourfold geometry, there is a prescription to assign the necessary flux quanta for those models which can be described in terms of the spectral cover. In this context it was observed that the flux distribution is subject to certain constraints which are common to all consistent models, referred to as the Dudas--Palti (DP) relations \cite{Dudas:2010zb}. Later it was shown that these conditions are nothing but the requirement of anomaly cancellation for the various gauge factors in four dimensions\cite{Marsano:2010sq}. Given their origin, it is thus expected that the DP relations are automatically fulfilled in global constructions and hence one expects them to hold beyond spectral cover models. In this section we review the DP relations following the arguments given in \cite{Marsano:2010sq}.

After switching on fluxes, the net amount of chiral matter in the representation $\mathbf{R}$ from a matter curve $\Sigma$ is counted via the following index theorem
\begin{equation}
 \chi(\mathbf{R})=\int_{\Sigma} c_1\left(V_\Sigma\otimes L_Y^{Y_{\mathbf{R}}}\right)=\int_{\Sigma} \left[c_1(V_\Sigma)+{\rm rk}(V_\Sigma)\, c_1(L_Y^{Y_\mathbf{R}})\right],
\end{equation}
where the bundle $V_\Sigma$ accounts for the $G_4$ flux and $L_Y$ is a line bundle used to specify the hypercharge flux and $Y_\mathbf{R}$ denotes the hypercharge carried by the representation $\mathbf{R}$. The previous relation can be split as
\begin{equation}
\chi(\mathbf{R})=\underbrace{\int_{\Sigma} c_1(V_\Sigma)}_{\mathcal{M}_\Sigma}+ Y_\mathbf{R} \underbrace{\left[{\rm rk}(V_\Sigma) \int_{\Sigma} \omega_Y\right]}_{\mathcal{N}_\Sigma}\, ,\label{chir}
\end{equation}
where we introduced the $(1,1)$-form $\omega_Y \sim c_1(L_Y)$ which is trivial in the full Calabi--Yau, as needed in order to ensure that the hypercharge remains massless \cite{Beasley:2008kw,Donagi:2008kj}. The quantities $\mathcal{M}_\Sigma$ and $\mathcal{N}_\Sigma$ are the same for all $\mathbf{R}$'s that belong to the same curve. With this we arrive at the following chiralities for the SM components originating from a $\mathbf{10}$- or a $\overline{\mathbf{5}}$-curve
{\small
\begin{equation}
\label{eq:MatterSplitting}
\begin{array}{lllllllll}
\Sigma_{\mathbf{10}_a}: & \, \,  (\mathbf{3},\mathbf{2})_{1/6} & : & M_a\,,  & \, \, \, \, \, \, \, \, & \Sigma_{\overline{\mathbf{5}}_{i}}: & \, \, (\overline{\mathbf{3}},\mathbf{1})_{1/3}& : & M_i\,,\\
 & \, \,  (\overline{\mathbf{3}},\mathbf{1})_{-2/3} & : & M_a-N_a\,,  & \, \, \, \, &  & \, \, (\mathbf{1},\mathbf{2})_{-1/2} & : & M_i+N_i\,,\\
& \, \,  (\mathbf{1},\mathbf{1})_{1} & : & M_a+N_a\,,  & \, \, \, \, &  & & \,& \,
\end{array}
\end{equation}}
where the non-curly $M$'s and $N$'s are related to the quantities in \eqref{chir} via
\begin{align}
\begin{split}
M_a&=\mathcal{M}_a+\frac{1}{6}\mathcal{N}_a\,, \qquad M_i=\mathcal{M}_i+\frac{1}{3}\mathcal{N}_i\,,\\
N_a&=\frac{5}{6}\mathcal{N}_a\,, \quad\qquad\qquad N_i=-\frac{5}{6}\mathcal{N}_i\,.
\end{split}
\end{align}

Having obtained the expressions for the field multiplicities, we can now discuss the anomaly cancellation conditions in generic F-theory models with the gauge group $\SU5\times\U1^N$. First of all, from the vanishing of anomalies including SM gauge factors only, one obtains
\begin{equation}
\sum_i M_i-\sum_a M_a=0\,,\label{zero}
\end{equation}
\begin{equation}
\sum_i N_i = \sum_a N_a = 0\, .\label{first}
\end{equation}
Note that the the first condition is reminiscent of the D7 tadpole cancellation condition in type~IIB models.

Mixed anomalies between the SM gauge factors and the \U1 symmetries outside of \SU5 do not need to vanish. However, the axionic shift which cancels the anomaly at the \SU5 level does not get modified after GUT breaking, because the hypercharge flux does not couple to the closed string sector \cite{Palti:2012dd}. This requirement leads to the following condition on the flux quanta
\begin{equation}
\sum_aq_a^\alpha N_a+\sum_iq_i^\alpha N_i =0\,, \label{second}
\end{equation}
where $q_a^\alpha$ and $q_i^\alpha$ are the corresponding charges for the \ten- and \fiveb-curves of the additional ${\rm U(1)}_\alpha$ symmetries.

An additional condition arises from anomalies of the form $\U1_Y$--${\rm U(1)}_\alpha$--${\rm U(1)}_\beta$, which would vanish as they descend from the trivial ${\rm SU(5)}$--${\rm U(1)}_\alpha$--${\rm U(1)}_\beta$ anomalies~\cite{Palti:2012dd}. Such conditions read
\begin{equation}
3\sum_aq^\alpha_a q^\beta_a N_a+\sum_iq^\alpha_i q^\beta_i N_i=0\, .\label{third}
\end{equation}
However, in contrast to \eqref{first} and \eqref{second}, this condition does not have any counterpart in terms of homology constraints obtained from spectral cover considerations. In particular, it has been shown in \cite{Palti:2012dd} that only very few of those constructions are in agreement with \eqref{third}. Interestingly, it was shown in an exploration of perturbative type IIB models that the so-called orientifold-odd GS mechanism could serve to cancel some of these anomalies \cite{Mayrhofer:2013ara}. Nonetheless, the `F-theory version' of this mechanism is not known yet, so it is still unclear whether \eqref{third} has to be imposed as a necessary constraint for (semi-)local model building in F-theory.

\subsection{Status of spectral cover model building}
\label{sec:scmodelbuilding}
As mentioned above the spectral cover constructions yield an explicit framework for realistic model building attempts within F-theory~\cite{Dudas:2009hu,Ludeling:2011en,King:2010mq,Marsano:2009wr,Marsano:2009gv,Marsano:2012yc}. However, when building GUT models within F-theory one has to deal with similar problems as those faced in standard SUSY GUTs. First, one needs to ensure that the triplets accompanying the Higgs multiplets are decoupled from the low energy theory. This is achieved by breaking the GUT group via hypercharge flux; with this flux it is possible to project out the triplets in the Higgs multiplets. Second, one needs to guarantee that all couplings are well under control, such that, for example dangerous operators which mediate fast proton decay are sufficiently suppressed. For this purpose additional \U1 symmetries are used as they appear naturally in this framework.

Let us start the discussion on F-theory model building by introducing the MSSM superpotential at the level of \SU5 up to dimension five
\begin{align}
\begin{split}
 \mathcal{W}=&\,\,\,\mu \five_{H_u} \fiveb_{H_d}+\beta_i \fiveb_{i} \five_{H_u}  \\
&+Y^u_{ij} \ten_i \ten_j \five_{H_u} + Y^d_{ij} \fiveb_i \ten_j \fiveb_{H_d} + W_{ij} \fiveb_i \fiveb_j \five_{H_u} \five_{H_u}\\
&+\lambda_{ijk} \fiveb_{i} \fiveb_{j} \ten_{k}+\delta_{ijk} \ten_{i} \ten_{j} \ten_{k} \fiveb_{H_d}+ \gamma_i \fiveb_{i} \fiveb_{H_d} \five_{H_u} \five_{H_u}\\
&+\omega_{ijkl}\ten_{i} \ten_{j} \ten_{k} \fiveb_{l}\, ,\label{sup}
\end{split}
\end{align}
in which the representations $\fiveb_{i}$ and $\ten_{i}$ correspond to the $i$-th family and $\five_{H_u}$, $\fiveb_{H_d}$ are the \SU5 multiplets giving rise to the up- and down-type Higgs respectively. The operators in the first line of \eqref{sup} are those leading to the $\mu$-term and the bilinears between $H_u$ and the lepton doublets. In the second line we have the Yukawa couplings $Y^{u,d}_{ij}$ and the Weinberg operator $W_{ij}$. R-parity violating dimension four and five operators are given in the third line, whereas the R-parity allowed dimension five operator (that leads to proton decay) is given in the fourth line. One also has dangerous proton decay operators arising from the K\"{a}hler potential~\cite{Barbier:2004ez,Dudas:2009hu}, namely
\begin{align}
\label{kaehler}
\mathcal{K}\supset \kappa_{ijk} \ten_{i} \ten_{j} \five_{k}+ \overline{\kappa}_i \fiveb_{H_u}\fiveb_{H_d}\ten_{i}  \,.
\end{align}

The models can be relevant for phenomenology if all dangerous operators and the $\mu$-term in \eqref{sup} and \eqref{kaehler} are forbidden by virtue of some \U1 symmetries.
 It is also desired that the top-quark Yukawa coupling is generated at tree level (i.e.~allowed by the \U1 symmetries). 
These U(1) symmetries are typically GS massive but are present as global symmetries in the effective field theory (EFT). However, often these U(1) symmetries can be broken for example by singlet VEVs or instanton effects~\cite{Cvetic:2011gp,Kerstan:2012cy,Marsano:2008py}. Independently of the breaking mechanism, discrete subgroups can remain unbroken and hence some operators can still be protected in the EFT. The crucial property is the difference in the U(1) charges for desired and undesired couplings. The breakdown of these U(1) symmetries has been used to realize a Froggatt--Nielsen (FN) type explanation of the flavor structure in the quark and lepton sector \cite{Dudas:2009hu,Dolan:2011iu}.

Previous searches for models in the spectral cover constructions have been based on the idea that the three families arise from complete \SU5 representations (i.e.\ from curves on which the hypercharge flux acts trivially). In addition to that one has two further $\five$-curves on which the hypercharge flux acts by projecting out the triplet components so that one ends up with only one pair of doublets (namely the Higgses). The couplings of the fields are determined by the extra \U1 symmetries and additional singlet fields in the spirit of FN, as described in the previous paragraph.

After choosing the factorization, the matter representations with their corresponding \U1 charges are fixed, and the model building is based on tuning the flux quanta carried by the different curves. This is not arbitrary and one has to ensure that the anomaly cancellation conditions \eqref{first}-\eqref{second} are satisfied. It turns out that these restrictions are far from trivial and severely constrain the possibilities for promising models. Previous explorations have lead to the following two observations:
\begin{itemize}
 \item If one insists on the exact MSSM spectrum, there is only one flavor-blind \U1 symmetry available. This symmetry corresponds to a linear combination of hypercharge and $\U1_{B-L}$. As it allows for a $\mu$-term as well as dimension five proton decay \cite{Dolan:2011iu}, this construction is not suitable for phenomenology.
 \item If one requires the presence of a \U1 symmetry which explicitly forbids the $\mu$-term, i.e.~a so-called Peccei--Quinn (PQ) symmetry, this implies the existence of exotic fields which are vector-like under the Standard Model gauge group and come as incomplete representations of the underlying \SU5 \cite{Dudas:2010zb}.
\end{itemize}
These observations suggest a strong tension between a solution to the $\mu$-problem and the absence of light exotics in the spectrum. As already mentioned, in these models the matter arise from \SU5 representations which are not split by hypercharge flux. Here we explore whether in models with split multiplets these tensions can be avoided and whether, in principle, these permit us to obtain an exotic free spectrum with appealing operator structure.

An additional phenomenological constraint is the requirement of correct gauge couplings at the observed energy scales which in the context of SUSY GUTs is translated into the question whether gauge coupling unification is consistent with low-energy data. To address this question, it should be noted that not only the matter content below the unification scale but also high-scale threshold effects influence unification. In particular, it was argued that some high-scale threshold effects can appear and can spoil unification but can also be absent depending on the exact geometry and flux background\cite{Donagi:2008kj,Blumenhagen:2008aw}. The combined effect of such potential threshold corrections and exotics has been explored in \cite{Dolan:2011aq}. As the exact presence of these high-scale threshold corrections depends on the structure of the geometric background, their exact size has to be calculated in the context of the new backgrounds with rational sections. However, note that the inclusion of split-multiplets (i.e. MSSM matter from incomplete \SU5 multiplets) is expected to add an additional effect to the high-scale threshold corrections as explicitly analyzed in the context of heterotic orbifold compactifications \cite{Ross:2004mi}. We do not address the question of unification at this state as it is beyond the scope of this paper, and assume for now that in F-theory, a model with the exact MSSM spectrum can comply with grand unification.

\section{Systematic F-theory model building\label{sec:ModelBuilding}}
In this section we aim at constructing models with the exact MSSM spectrum (plus singlet extensions) based on \SU5 F-theory GUTs with additional \U1 symmetries emerging from spectral cover or rational section constructions. In this spirit, we first discuss the consistency conditions and phenomenological constraints we implement along our search. After that we present the results for spectral cover and rational section models. Finally we apply the same criteria in order to find bottom-up models which share the same overall features regarding the \U1 charges but go beyond the explicit rational section models considered to date.

\subsection{Search strategy}
\label{sec:Search}
The first requirement of our search is that the models have the MSSM matter content (i.e.~three families of quarks and leptons and one pair of Higgses) up to possible SM singlets and extra \U1 gauge bosons. Thus, from the beginning we impose the absence of extra states charged under the Standard Model gauge group. We also require the spectrum to satisfy the four-dimensional anomaly cancellation conditions \eqref{first}-\eqref{second} such that the masslessness of the hypercharge is guaranteed. With regards to this, we should remark that in the class of models we study it is not possible to make the anomalies of the type $\U1_Y$--$\U1_\alpha$--$\U1_\beta$ to comply with condition  \eqref{third} (even in the cases where one allows for exotic matter). For this reason we have to rely on a special type of GS mechanism to cancel this anomaly. As mentioned in Section~\ref{Sec:DP}, such a mechanism has been observed in the specific context of perturbative type IIB string theory but its F-theory uplift is not known yet. We also explore the possibility of having phenomenologically appealing models which comply with the condition \eqref{third} together with the other anomaly requirements. For this purpose we implement a bottom-up survey with at most two \U1 symmetries in addition to the \SU5 GUT group. The charges for the fields under these \U1 symmetries are set to fit a particular splitting as given in Table \ref{tab:Charges}. We elaborate more on this survey in Section \ref{sec:Alternatives}.
\medskip

The phenomenological requirements we impose on our models at tree level are the presence of a top quark Yukawa coupling and the absence of operators that induce proton decay and the $\mu$-term. We then look at the \U1 charges carried by all relevant operators in order to unveil the mechanisms which could generate all desired operators (e.g.~Yukawa couplings) while keeping dangerous couplings well under control. Among those mechanisms one could think of the Froggatt-Nielsen mechanism in which some singlets get a VEV to induce certain effective couplings, but instantonic or other effects could also serve for this purpose.

For both the spectral cover and the rational sections models we scan over possible choices for the hypercharge and chirality flux. This input determines the spectrum and the allowed couplings, and we check which couplings satisfy the aforementioned constraints. Note that we do not explicitly require complete \SU5 matter multiplets after turning on fluxes. 

\subsubsection*{Matter content and anomaly cancellation conditions}
After fixing the matter curves and their \U1 charges from a given spectral cover or rational section model, the only freedom left is to switch on the fluxes such that the desired MSSM content is obtained and the four-dimensional anomaly cancellation conditions are satisfied. Using the notation of Section~\ref{Sec:DP}, we obtain the following requirements for the flux choices:\\
Requiring {\bf three chiral families} imposes that the chirality flux has to satisfy
\begin{align}
\sum_{\Sigma_{\ten}} M^a=\sum_{\Sigma_{\fiveb}} M^i= 3 \quad \text{ with } \quad M^a\,,\, M^i \geq 0\,.
\end{align}
This relation guarantees that the anomaly constrain from \eqref{zero} is satisfied.
We demand that in addition to the MSSM matter content, {\bf no exotics} are present in the spectrum (with the exception of singlet fields neutral under the SM gauge group). This requirement constrains the quanta of hypercharge flux to obey the following relations
\begin{align}
\sum_{\Sigma_{\ten}} N^a =& 0 \quad \text{ with }\quad  -M^a \leq N^a \leq M^a\,  ,\\
\sum_{\Sigma_{\fiveb}} N^i =& 0 \quad \text{ with }\quad  -M^i-1 \leq N^i \leq 3\, .
\end{align}
Note that this flux configuration automatically satisfies the constraints \eqref{first}. The additional flux constraint on the \five-curves
\begin{align}
\sum_{\Sigma_{\fiveb}} |M^i + N^i| =5\, ,
\end{align}
guarantees three lepton doublets together with {\bf exactly one pair of Higgses}.

The above constraints fix the matter spectrum to that of the MSSM; note however that additional singlets under the SM gauge group are typically present. For a configuration with such a spectrum, one has to check whether the flux choices satisfy the additional constraints \eqref{second} from anomaly cancellation. As already mentioned, at this point we do not take the condition \eqref{third} into account.

\subsubsection*{Phenomenological constraints}
We also have to constrain the operators of the effective field theory in order to obtain a realistic model. As the charges for the fields are given ab initio, we have to ensure that dangerous operators are not generated at tree level, so that the \U1 charges for these operators have to be non-zero. Since we are not dealing with complete \SU5 multiplets, it is necessary to decompose the \SU5 couplings \eqref{sup} and \eqref{kaehler} in terms of the SM fields.
In order to have a heavy top quark we demand the presence of the top Yukawa coupling at tree level, i.e.\ we require that at least one of the up-type Yukawa couplings
\begin{align}
\ten_i \ten_j \five_{H_u} \supset Q_i \bar{u}_j H_u
\end{align}
is allowed by all \U1 symmetries. In the case where all $Q_i$ and $\bar{u}_j$ descend from only one \ten-curve the up-quark Yukawa matrix is of rank one. Nevertheless, this matrix can acquire full rank when appropriate flux or non-commutative deformations~\cite{Heckman:2008qa,Cecotti:2009zf,Hayashi:2009ge,Hayashi:2009bt,Conlon:2009qq,Font:2013ida,Font:2012wq} away from the ${\rm E}_6$ Yukawa point are included.

In order for low energy SUSY to {\bf solve the $\mu$-problem}, we require the $\mathbf{\mu}$-term
\begin{align}
 \mu \five_{H_u} \fiveb_{H_d} \supset  \mu H_d H_u
\end{align}
to be forbidden by any of the \U1 symmetries. The above coupling will be generated upon breakdown of these symmetries. Let us also remark that in addition to the expected suppressions in the couplings due to singlet VEVs (or instantons) some additional suppression is expected when the couplings arise from the K\"{a}hler potential. This is fact is particularly appealing for the generation of the $\mu$-term as it can be sufficiently small, if induced from the K\"{a}hler potential along the lines of the so-called Giudice-Masiero (GM) mechanism~\cite{Giudice:1988yz}.
 
The $\mu$-term is closely linked to the presence of dimension five $B$-$L$ invariant operators (see Section~\ref{sec:ResultsRS} for further discussions)
\begin{align}
\omega_{ijkl}\ten_{i} \ten_{j} \ten_{k} \fiveb_{l} \supset \omega^{1}_{ijkl}Q_i Q_j Q_k L_l + \omega^{2}_{ijkl}\bar{u}_i \bar{u}_j \bar{e}_k \bar{d}_l +\omega^{3}_{ijkl}Q_i \bar{u}_j \bar{e}_k L_l  \, .
\end{align}
We demand these operators to be forbidden by the \U1 charges, keeping in mind that they can be generated in a similar fashion as the $\mu$-term. 

In order to {\bf avoid fast proton decay}, the \U1 symmetries must also forbid the following superpotential and K\"{a}hler potential couplings:
\begin{align}
\label{eq:dangerousCouplings}
\begin{array}{l@{}l@{\;}l}
\beta_i & \fiveb_{i} \five_{H_u} &\supset \beta_i L_{i}H_u    \, ,\\
\lambda_{ijk}& \fiveb_{i} \fiveb_{j} \ten_{k} &\supset  \lambda^{0}_{ijk} L_i L_j \bar{e}_k + \lambda^{1}_{ijk}  \bar{d}_i L_j Q_k   + \lambda^{2}_{ijk}  \bar{d}_i \bar{d}_j \bar{u}_k  \, ,\\
\delta_{ijk}& \ten_{i} \ten_{j} \ten_{k} \fiveb_{H_d} &\supset \delta^1_{ijk} Q_i Q_j Q_k H_d +  \delta^1_{ijk} Q_i \bar{u}_j \bar{e}_k H_d  \, ,\\
\gamma_i& \fiveb_{i} \fiveb_{H_d} \five_{H_u} \five_{H_u} &\supset \gamma_i L_i H_d H_u H_u  \, ,\\
\kappa_{ijk}& \ten_{i} \ten_{j} \five_{k} &\supset \kappa^{1}_{ijk}Q_i \bar{u}_j \bar{L}_k + \kappa^{2}_{ijk} \bar{e}_i \bar{u}_j d_k + \kappa^{3}_{ijk} Q_i Q_j d_k  \, ,\\
\overline{\kappa}_{i}& \fiveb_{H_u}\fiveb_{H_d}\ten_{i} &\supset \overline{\kappa}^{1}_i H^*_u H_d \bar{e}_i \, .
\end{array}
\end{align}
For a consistent model we need to require that upon breakdown of the \U1 symmetries these operators are not generated. This is for example achieved by demanding the presence of an effective matter parity symmetry.

We also expect that while the operators in \eqref{eq:dangerousCouplings} remain absent, it is possible to generate {\bf full rank Yukawa matrices}\footnote{Recall that, as the matter fields need not to arise from complete \SU5 multiplets, so that the down and lepton Yukawas do not necessarily coincide.} $Y^u_{ij}$, $Y^d_{ij}$ and $Y^L_{ij}$. This necessarily implies that the charges of the desired operators must differ in comparison to the undesired ones. As a consequence, one observes that the field $H_d$ has to come from a different curve than all the other leptons and triplets to guarantee that dimension four operators (such as $\lambda^0$ and $\lambda^1$ in \eqref{eq:dangerousCouplings}) are not introduced together with the Yukawa entries.\footnote{We thank H.P.\ Nilles for pointing this out.}

\subsection{Models from spectral cover constructions}
\label{sec:SpectralC}
Among the spectral cover models we considered those which have at most two additional \U1 symmetries, corresponding to the splittings presented in Section~\ref{sec:SpectralCoverReview}. The requirement of having $H_d$ and no further leptons arising from the $H_d$ curve, together with the requirement of forbidding all dangerous operators at tree level, leads to only six models with the MSSM spectrum, all of which are based on the 2+2+1 splitting. Unfortunately, all these models are phenomenologically unappealing for various reasons. We present one model to exemplify the generic problems of those setups.
\begin{table}[t]
\centering
\renewcommand{\arraystretch}{1}
\begin{tabular}{|c|c|c|cc|c|}
\hline
 Curve & $q_1$ & $q_2$ & $M$ & $N$ & Matter\\
\hline
$\mathbf{10}_{1}$ & $1$ & $5$ & 2 & 1 & $Q_{1,2}+\bar{u}_{1}+\bar{e}_{1,2,3}$\\
$\mathbf{10}_{5}$ & $-4$ & $0$ & 1 & -1 & $Q_{3}+\bar{u}_{2,3}$\\
$\overline{\mathbf{5}}_{11}$ & $2$ & $10$ & 0 & -1 & $H_u^c$\\
$\overline{\mathbf{5}}_{15}$ & $-3$ & $5$ & 0 & 1 & $H_d$ \\
$\overline{\mathbf{5}}_{35}$ & $-3$ & $-5$ & 3 & 0 & $(L+\bar{d})_{1,2,3}$\\
\hline
\end{tabular}
\caption{Spectral cover model and its corresponding flux quanta along the different matter curves in the 2+2+1 splitting. The lower indices of the matter representations are family indices.}
\label{tab:M2}
\end{table}
The matter content and the \U1 charges are summarized in Table~\ref{tab:M2}. The main problem of these models lies in the tension for the charges among the following operators:
\begin{equation}
q(Q_2 \bar{u}_2 H_u)=q(H_u H_d)=q(Q_1 \bar{d}_1 L_i)=q(\bar{u}_1  \bar{u}_1 \bar{d}_k)=(-5,-5)\, .
\end{equation}
There is, at this level, no mechanism available to generate a hierarchy among these operators, hence in this class of models we would expect some dimension four $B$-$L$ violating operators to be of the same order as some Yukawa couplings and this is an undesired feature in a realistic model.

\subsection{Constructions with rational sections}
\label{sec:ResultsRS}
In this section we repeat our model building analysis for models that are derived from the rational section constructions introduced in Section~\ref{sec:ToricSectionsReview}. We restrict ourselves to the subset of models whose tops have already been examined in the literature. Compared to the spectral cover models, they have the advantage that they allow for more general \U1 charge assignments, but the restriction that all \ten matter curves have the same U(1) charges\footnote{Note that this restriction is true for all models considered here but is not a generic feature of rational section models.}. Hence we cannot put hypercharge flux along those. This restricted setup also allows us to find some analytic relations between the \U1 charges of certain operators which substantially influence the phenomenological properties of the models.

\subsubsection*{\U1 Charge pattern}

In the context of rational section models, the requirements outlined in Section~\ref{sec:Search} simplify due to the fact that all \ten-curves have the same charge, which implies that all \ten-plets stay complete. For that reason, we suppress the family indices of their corresponding Standard Model representations.

First of all, the presence of a tree level top Yukawa fixes the $H_u$ charge to be
\begin{align}
q(H_u) = - 2q(\mathbf{10})\,.
\end{align}

For the subsequent discussion, we introduce the following notation for the charges of the operators:
\begin{align}
\begin{split}
\mu\,:& ~~ q(H_d H_u)\,=q(H_d)+q(H_u):=q^{\mu} \,, \\
Y^L\,:& ~~ q(\bar{e} H_d L_i):= q^{Y^L_i} \,, \\
Y^d\,:& ~~ q(\bar{u} H_d \bar{d}_i)_{\,}:=q^{Y^{d}_i} \, , \\
\beta_i:& ~~ q(L_i H_u) ~\,= q(L_i)+q(H_u) := q^{\beta_{j}}\, .
\end{split}
\end{align}
Among these operators, all but the $\beta_i$ terms should be induced upon breakdown of the \U1 symmetries.

Now we can express the charges of all unwanted operators in terms of the charges defined above. The dangerous dimension four proton decay operators are:
\begin{align}
\label{noT}
\begin{split}
\lambda^0_{ij}& \,:~~ q(Q\bar{d}_i L_j) = q^{Y^d_i}+q(H_d)-q(L_j) =q^{Y^d_i}+ q^\mu - q^{\beta_j}\\
\lambda^1_{ij}& \,:~~ q(\bar{e} L_i L_j)_{\,} =q^{Y^L_i}+ q(H_d)-q(L_j)=q^{Y^L_i}+ q^\mu - q^{\beta_j} \\
\lambda^2_{ij}& \,:~~ q(\bar{u} \bar{d}_i \bar{d}_j)\;= q^{Y^d_i} + q(H_d) - q(\bar{d}_j)~ = q^{Y^d_i}+ q^\mu -q(H_u) - q(\bar{d}_j) \, .
\end{split}
\end{align}
Since we want to generate the down-type Yukawa matrices, we see that the previous couplings are only forbidden due to the charge difference between the $H_d\,$- and $L_j\,$-curves in the case of the $\lambda^0_{ij}$ and $\lambda^1_{ij}$ couplings, and due to the charge difference between $H_d\,$- and $\bar{d}_j\,$-curves in the case of the $\lambda^2_{ij}$. Thus, as already pointed out, it is necessary that the $\five_{H_d}\,$-curve contains only the down-type Higgs, since any lepton or down-type quark with identical charge will automatically induce a dangerous operator. As we want to induce the $\mu$-term as well, we observe that no $\bar{d}_i$ field can arise from the $H_u$-curve either.\footnote{Note that if $H_u$ and $\bar{d}_j$ come from the same curve their \U1 charges carry opposite signs.}

Overall, note that the charges can also be written in terms of those of the forbidden operators $\beta_i$. Thus, if we find a configuration  such that the Yukawa couplings and the $\mu$-term is induced but the $\beta_i$-terms stay forbidden, the dimension four operators stay forbidden as well.

Furthermore, we observe that the dimension five operators in the superpotential
\begin{align}
\begin{split}
\omega^1_i\,,\omega^3_i&\, :~~ q(Q Q Q L_i) = q(Q\bar{u}\bar{e} L_i)= - q^{\mu}+ q^{Y^L_i}:= q(\mathbf{10}\,\mathbf{10}\,\mathbf{10}\, L_i)  \, ,\\
\omega^2_i& \,:~~ q(Q Q \bar{u} \bar{d}_i   )\,\, = \,q(\bar{u}\bar{u}\bar{e}\bar{d}_i)\;= - q^{\mu}+ q^{Y^d_i}:= q(\mathbf{10}\,\mathbf{10}\,\mathbf{10}\, \bar{d}_i) \, ,
\end{split}
\end{align}
will be unavoidably induced together with the Yukawa couplings and the $\mu$-term. It should be noted that the $\mu$-term charge enters with a minus sign in the previous equations. This implies that the mechanism (such as a singlet VEV) which induces the $\omega^i$-terms in the superpotential will not induce the $\mu$-term directly in the superpotential but can generate it from the K\"{a}hler potential after SUSY breakdown. Estimating the exact suppression of the resulting coupling seems very interesting but is beyond the scope of this paper. Note also that it is possible to induce a Weinberg operator
\begin{align}
W_{ij}:~~ q(L_i L_j H_u H_u) &=q^{\beta_i} + q^{\beta_j}
\end{align}
without inducing the $\beta_i$-terms by using, for example, singlet VEVs with charge $q(s_i)=-2q^{\beta_i}$.

In a similar fashion, we observe that the operators
\begin{align}
\begin{split}
\delta^1 \,, \delta^2:&~~ ~\;q(Q Q Q H_d)\,~ = q(Q\bar{u}\bar{e} H_d) = - q^{\beta_i}+ q^{Y^L_i} \, ,\\
\gamma_i:&~~ q(L_i H_d H_u H_u)= q^{\mu} + q^{\beta_i} \, ,
\end{split}
\end{align}
will remain absent as long as the $\beta_i$-terms are not induced. The same holds for the K\"{a}hler potential terms
\begin{align}
\begin{split}
\kappa^1_i:& ~~ \;q(Q\bar{u} L^*_i)\; = -q^{\beta_i}\, , \\
\overline{\kappa}:& ~~ q(\bar{e}H^*_u H_d )= q^{\mu} + q^{\beta_i}\, ,
\end{split}
\end{align}
with the exception of
\begin{align}
\kappa^2_i\,,\kappa^3_i:& ~~ q(QQ \bar{d}^*_i) = q(\bar{u} \bar{e} \bar{d}^*_i) = -q(H_u)-q(\bar{d}_j)=-q^\mu + q(H_d) - q(\bar{d}_i):=q(\mathbf{10}\,  \mathbf{10}\, \bar{d}^*_i) \, , \label{noT2}
\end{align}
for which one has to ensure that no triplets emerge from the Higgs curves as a necessary (but not sufficient) condition.

Note that the above observations are independent of the number of \five-curves and \U1~symmetries. However, there remains a crucial interplay between the Higgs charges compared to those of the down-type quarks and those of the singlet fields, which have to be checked on a case by case analysis.

\subsubsection*{Results of the scan}
Based on the previous considerations, we have scanned over all available fiber polygons. We find that models with a single \U1 factor do not allow for a configuration with the desired properties. This rules out the polygons $F_3$, $F_8$ and $F_{11}$ of \cite{Braun:2013nqa} as potential candidates\footnote{It could be that certain choices of the base allow for further non-toric sections. In that case one could expect models from such fibers.} and leaves us with $F_5$, which allows for two toric \U1 symmetries, and its corresponding four inequivalent \SU5 tops \cite{Borchmann:2013jwa}.

\renewcommand{\arraystretch}{1.2}
\begin{table}[t]
\centering
\begin{tabular}{|c||cccccc|}\cline{1-7}
\multicolumn{1}{|c||}{top} & MSSM fields & No anoms & Heavy top & $q^\lambda \neq q^{Y}$ & $p$ stable &  suitable VEV \\ \hline
1 & 5795 & 140 & 26 & 18 & 4 & 2 \\
2 & 5795 & 140 & 27 & 22 & 3 & 2 \\
3 & 5795 & 140 & 34 & 29 & 0 & 0 \\
4 & 5795 & 140 & 27 & 22 & 0 & 0 \\
\hline
\end{tabular}
\caption{Results of the model scan with two \U1 symmetries using rational sections. Every step in our search strategy reduces the amount of models, leaving four satisfactory models in the end, out of which two are inequivalent.}
\label{tab:RSStatistics}
\end{table}
\renewcommand{\arraystretch}{1.0}

In Table~\ref{tab:RSStatistics} we summarize the details of the scan for spectral cover and rational section models. We want to emphasize that we start from $\mathcal{O}(2\cdot 10^4)$ rational section models that feature the exact MSSM matter content but only four of them feature a viable phenomenology. For every top one obtains 5795 models with the spectrum of the MSSM. The fact that up to this point all tops feature the same amount of models is due to the similar charge pattern they exhibit, giving rise to similar flux solutions. From the second column in Table~\ref{tab:RSStatistics} we see that the anomaly constraints \eqref{second} reduce the available models considerably. In all cases, the requirement of a tree level Yukawa for the top further reduces the amount of models by about $80$ percent, cf.\ column three. In the end only the first two tops provide models in which all undesirable operators are forbidden by the \U1 symmetries. We find four models that satisfy all conditions discussed in Section~\ref{sec:Search}. Out of these, only two feature different phenomenology. We present one of these benchmark models in the following, while the other one can be found in Appendix \ref{sec:AppendixBenchmarkModel2}.

\begin{table}
\centering
\renewcommand{\arraystretch}{1.4}
\centering
{\footnotesize
\begin{equation*}
\begin{array}{|c|cc|cc|c|cc}
\cline{1-6}\cline{8-8}
\multicolumn{6}{l}{\text{\bf 1. Spectrum }} & & \multicolumn{1}{l}{\text{\bf 2. Singlet VEVs:}\,\, \boldsymbol{s_1\,, a}}\\
\cline{1-6}\cline{8-8}
\multicolumn{6}{l}{} & & \multirow{3}{*}{$q(s_1)=(0,5)\,,\quad q(a)=(10,0)$\,.}\\
\cline{1-6}
\text{Curve} & q_1 & q_2 & M & N & \text{Matter} & & \\ \cline{1-6}
\ten   & -1& 2& 3 & 0 & (Q+\bar{u}+\bar{e})_{1,2,3} & &\\
\cline{8-8}
\fiveb_1 & 3 & -1& 1 & -1& \bar{d}_1& & \multicolumn{1}{l}{\text{\bf 3. $\boldsymbol{\mu}$- and $\boldsymbol{\beta_i}$-terms}}\\
\cline{8-8}
\fiveb_2 & -2 & 4& 0 & -1& H_u & & \multirow{3}{*}{$q(H_u \bar{L}_i)=(5,0)\, ,\quad q(H_u H_d) = (0,-5)\,.$}\\
\fiveb_4 & 3 & 4 & 2 & 1 & L_{1,2,3}+\bar{d}_{2,3} & & \\
\fiveb_5 & -2& -1& 0 & 1 & H_d & &\\
\cline{1-6}
\multicolumn{6}{l}{} & &  \\
\hline
\multicolumn{8}{l}{\text{\bf 4. Yukawa couplings}} \\
\hline
\multicolumn{8}{c}{} \\
\multicolumn{8}{c}{q(Q_i\bar{u}_j H_u) =(0,0)\, ,\quad\quad q(Q_i\bar{d}_j H_d) =
\begin{pmatrix}
 (0,0) \\
 (0,5) \\
 (0,5)
\end{pmatrix}_j \, ,\quad\quad q(\bar{e}_i L_j H_d)=(0,5)\,.} \\
\multicolumn{8}{c}{} \\
\hline
\multicolumn{8}{l}{\text{\bf 5. Allowed dimension five proton decay and Weinberg operators}} \\
\hline
\multicolumn{8}{c}{} \\
\multicolumn{8}{c}{q(\ten\,\ten\,\ten\, L_i) = (0,10) \, , \quad\quad q(\ten\,\ten\,\ten\,\bar{d}_i)= \begin{pmatrix}
(0,5) \\
(0,10)\\
(0,10)
\end{pmatrix}_j\,,\quad\quad q(L_i\,L_j\,H_u\, H_u)= (10,0)\,.} \\
\multicolumn{8}{c}{} \\
\hline
\multicolumn{8}{l}{\text{\bf 6. Forbidden operators}} \\
\hline
\multicolumn{8}{c}{} \\
\multicolumn{8}{c}{q(\bar{u}\bar{d}_i\bar{d}_j)=
\begin{pmatrix}
(5,0)&(5,0)&(5,0) \\
(5,0)& (5,10) & (5,10) \\
(5,0)& (5,10) & (5,10)
\end{pmatrix}_{i,j}
 \, ,\quad\quad q(\ten\,\ten\, \bar{d}^*_i)= \begin{pmatrix} (-5,5) \\ (-5,0) \\ (-5,0) \end{pmatrix}\,.}
\end{array}
\end{equation*}
}
\caption{Details of benchmark model A. We give the charges for the operators $\lambda^2_{ij}$ and $\kappa^3_{ij}$, $\kappa^2_{ij}$ discussed in \eqref{noT} and \eqref{noT2}.}
\label{tab:BenchmarkA}
\end{table}

The properties of the model are shown in Table~\ref{tab:BenchmarkA}, where we have included the spectrum together with the singlet fields needed to generate the required couplings. We also give the charges of those dangerous operators which are automatically absent as long as no $\beta_i$-term is allowed. For this model we see that a singlet $s_1$ with charge $(0,5)$ that develops a VEV will generate a $\mu$-term from the K\"{a}hler potential and induce the Yukawa couplings and dimension five operators in the superpotential, while all dimension four operators stay forbidden. The orders of magnitude for these couplings are
\begin{align}
\begin{split}
 Y^L_{i}\sim\frac{\langle s_1 \rangle}{\Lambda}\,, &\hspace{1cm}Y^d_{i}\sim
\delta_{1,i}+\frac{\langle s_1\rangle}{\Lambda}(\delta_{2,i}+\delta_{3,i})\,,\\
\omega^1_i\,,\omega^3_i\sim \frac{\langle s_1\rangle^2}{\Lambda^3}\,,&\hspace{1cm}\omega^2_i\sim \frac{\langle s_1\rangle}{\Lambda^2}\delta_{1,i}+\frac{\langle s_1\rangle^2}{\Lambda^3}(\delta_{2,i}+\delta_{3,i})\,,
\end{split}
\end{align}
where $\Lambda$ is the appropriate cutoff scale, which depends on the global embedding of the local model. Following the analysis in~\cite{Hinchliffe:1992ad, BenHamo:1994bq}, the only severely constrained operators that are induced are 
\begin{align}
\label{eq:RSProtonDecayOperators}
\omega^1 \lesssim \frac{10^{-7}}{M_{\rm P}}\, \text{ and } \omega^2 \lesssim \frac{10^{-7}}{M_{\rm P}} \, .
\end{align}
All other operators that are induced stay unproblematic as long as no $\lambda^i$-terms are generated. The coupling $\omega^2$ only leads to a constraint if there is a non-diagonal degeneracy of quark and squark masses~\cite{Hinchliffe:1992ad, BenHamo:1994bq}. The $\omega^1$-operator, however, puts constraints on the size of the singlet VEV of $s_1$ that induces the operator after two insertions, to be
\begin{align}
\frac{\langle s_1\rangle^2}{\Lambda^3} &\lesssim \frac{10^{-7}}{M_{\rm P}} \, .
\end{align}
Such a size of the VEV seems compatible with down-quark and lepton-Yukawa couplings at the weak scale~\cite{Dreiner:2003hw}.

As mentioned before, it is possible to generate the Weinberg operator while keeping dangerous operators forbidden. In the matter spectrum of $F_5$
there exists no singlet with charge $(\pm10,0)$ whose VEV can introduce that operator. However, one might envision a non-pertur\-bative effect (e.g.~via instantons) which allows to generate this coupling. Note again that such an instanton resembles the effect of a singlet VEV and we will parameterize it by $\langle a \rangle$ such that the operator is introduced by\footnote{Note that the expected order of magnitude for this operator is the same for all generations as all lepton doublets in this model are found to arise from the same matter curve.}
\begin{align}
\label{eq:RSWeinbergOperator}
 W_{ij}\sim \frac{\langle a\rangle}{\Lambda^2}\,.
\end{align}

Another interesting question is which symmetries remain after the \U1 symmetries are broken. For this purpose it is more convenient to rotate the \U1 generators as specified in Table~\ref{tab:MatterParity}. There we see that after appropriate normalization, the charges of the singlets break the \U1 symmetries to a $\mathbbm{Z}_2$ subgroup under which the charges $q_1^\prime$ coincide with those of R-parity. On the other hand, we see that the second \U1 with charges $q_2^{\prime}$ gets broken completely by the VEV of $s_1$.
\begin{table}
\centering
\renewcommand{\arraystretch}{1.2}
\begin{center}
\begin{tabular}{|c|ccccc|cc|}
\cline{2-8}
\multicolumn{1}{c|}{}& $(Q+\bar{u}+\bar{e})_{1,2,3}$ &  $\bar{d}_1$ & $H_u$ & $L_{1,2,3}+\bar{d}_{2,3}$ &  $H_d$ & $s_1$ & $a$\\
\hline
$q_1$ & $-1$ & $3$ & $2$  & $3$ & $-2$ & $0$  & $-10$\\
$q_2$ & $2$  &$-1$ & $-4$ & $4$ & $-1$  & $-5$ & $0$\\
\hline
$q_1^{\prime}=(q_1-2q_2)/5$       & $-1$& $1$ & $2$ & $-1$ & $0$ & $2$ & $-2$\\
$q_2^{\prime}=(2q_1+q_1)/5$ & $0$ & $1$ & $0$ & $2$ & $-1$ & $-1$ & $-4$\\
\hline
\end{tabular}
\end{center}
\caption{The \U1 charges for the benchmark model in a rotated \U1 basis. Note that after giving VEVs, the charges $q_1^{\prime}$ are those of matter parity, whereas the charges $q_2^{\prime}$ become all trivial since $s_1$ has charge~$-1$.}
\label{tab:MatterParity}
\end{table}

\subsection{Beyond toric sections}
\label{sec:Alternatives}

In the previous section we only considered the subclass of polytopes which have been studied in the literature so far. For example, the polytopes $F_7$, $F_9$ and $F_{12}$ still remain to be analyzed. Also, one could have additional, non-toric, base dependent \U1 symmetries which could lead to different phenomenological patterns. As all of the above models do not satisfy the anomaly constraints \eqref{third} quadratic in the \U1 symmetries, one has to determine to which extent this constraint applies and how models can circumvent such constraints as for example suggested in the context of type IIB constructions~\cite{Mayrhofer:2013ara}.

In this section we want to explore possible bottom-up models in which the field theory anomaly coefficient actually vanishes so there is no need to rely on a geometric construction of canceling this anomaly. The aim is to provide a guideline for future geometric engineering efforts and it would be very interesting to obtain explicit realizations with such charges.

\begin{table}
\centering
\renewcommand{\arraystretch}{1.4}
\centering
{\footnotesize
\begin{equation*}
\begin{array}{|c|cc|cc|c|cc}
\cline{1-6}\cline{8-8}
\multicolumn{6}{l}{\text{\bf 1. Spectrum }} & & \multicolumn{1}{l}{\text{\bf 2. Singlet VEVs:}\,\, \boldsymbol{s_1\,, s_2}}\\
\cline{1-6}\cline{8-8}
\multicolumn{6}{l}{} & & \multirow{3}{*}{$q(s_1)=(0,\pm 5)\,,\quad q(s_2)=(\pm 10,0)$\,.}\\
\cline{1-6}
\text{Curve} & q_1 & q_2 & M & N & \text{Matter} & & \\ \cline{1-6}
\ten   & -3& -1& 3 & 0 & (Q+\bar{u}+\bar{e})_{1,2,3} & &\\
\cline{8-8}
\fiveb_1 & 9 & -2& 0 & 1& L_1& & \multicolumn{1}{l}{\text{\bf 3. $\boldsymbol{\mu}$- and $\boldsymbol{\beta_i}$-terms}}\\
\cline{8-8}
\fiveb_2 & 9 & -7& 1 & -1& \bar{d}_{1} & & \\
\fiveb_3 & -1 & 8 & 2 & -1 & L_2 + \bar{d}_{1,2} & & \multirow{3}{*}{\vspace{-2pt}$q(H_u L_i)=\begin{pmatrix}
 (15,0) \\
 (5,10) \\
 (5,-5)
\end{pmatrix}\, ,\quad q(H_u H_d) = (0,10)$\, .}\\
\fiveb_4 & -1& -7& 0 & 1 & L_3 & & \\
\fiveb_5 & -6& 8& 0 & 1 & H_d & &\\
\fiveb_6 & -6& -2& 0 & -1 & H_u & &\\
\cline{1-6}
\multicolumn{6}{l}{} & &  \\
\hline
\multicolumn{8}{l}{\text{\bf 4. Yukawa couplings}} \\
\hline
\multicolumn{8}{c}{} \\
\multicolumn{8}{c}{q(Q \bar{u} H_u) =(0,0)\, ,\quad\quad q(Q \bar{d}_j H_d) =
\begin{pmatrix}
 (0,0) \\
 (-10,15) \\
 (-10,15)
\end{pmatrix} \, ,\quad\quad q(\bar{e} L_j H_d)=\begin{pmatrix}
 (0,5) \\
 (-10,15) \\
 (-10,0)
\end{pmatrix}.} \\
\multicolumn{8}{c}{} \\
\hline
\multicolumn{8}{l}{\text{\bf 5. Allowed dimension five proton decay and Weinberg operators}} \\
\hline
\multicolumn{8}{c}{} \\
\multicolumn{8}{c}{q(\ten\, \ten\, \ten\, L_i) = \begin{pmatrix}
 (0,-5) \\
 (-10,5) \\
 (-10,-10)
\end{pmatrix} \, ,\,\quad q(\ten\, \ten\, \ten\, \bar{d}_i)= \begin{pmatrix}
(0,-10) \\
(-10,5)\\
(-10,5)
\end{pmatrix}\,,
} \\
\multicolumn{8}{c}{\qquad\qquad\qquad\, q(L_i\, L_j\, H_u\, H_u)= \begin{pmatrix}
(30,0)&(20,10)&(20,-5) \\
(20,10)& (10,20) & (10,5) \\
(20,-5)& (10,5) & (10,-10)
\end{pmatrix}\,.} \\
\multicolumn{8}{c}{} \\
\hline
\multicolumn{8}{l}{\text{\bf 6. Forbidden operators}} \\
\hline
\multicolumn{8}{c}{} \\
\multicolumn{8}{c}{q(\bar{u}\bar{d}_i\bar{d}_j)=
\begin{pmatrix}
(15,-15)&(5,0)&(5,0) \\
(5,0)& (-5,15) & (-5,15) \\
(5,0)& (-5,15) & (-5,15)
\end{pmatrix}
 \, ,\quad\quad q(\ten \ten \bar{d}^*_i)= \begin{pmatrix} (-15,5) \\ (-5,-10) \\ (-5,-10) \end{pmatrix}\,.}
\end{array}
\end{equation*}
}
\caption{Details of a benchmark model beyond the toric sections. We give the charges for the operators $\lambda^2_{ij}$ and $\kappa^3_{ij}$, $\kappa^2_{ij}$ discussed in \eqref{noT} and \eqref{noT2}.}
\label{tab:BenchmarkBeyondRS}
\end{table}

Based on the observation that the \U1 charges are all fixed (up to mod five) to a certain value given by the corresponding splitting and the intersection numbers of the fiber components with the rational section, we consider the possibility of having further models with two \U1 symmetries whose splitting gives rise to any of those charge assignments in Table~\ref{tab:Charges}. Furthermore, we assume that such models allow for more \five-curves, but only one \ten-curve as before. These additional \five-curves allow for more freedom to satisfy the anomaly constraints. The \five-curves are chosen to have charges
\begin{align}
q_{1,\fiveb} = Q_{1,\fiveb}+5\,n_{1,i}\,,\quad q_{2,\fiveb} = Q_{2,\fiveb}+5\,n_{2,i} \, ,
\end{align}
where $Q_{1,\fiveb}$ and $Q_{2,\fiveb}$ are fixed by the splitting that is chosen for each \U1. The integer valued $n_{1,i},n_{2,i}$ are in the range
\begin{align}
n_{1,i}\,,n_{2,i} \in \left[ -2,2\right]\,.
\end{align}
The charge of the \ten-curve is chosen such that it fits the structure of a given split, see Table~\ref{tab:Charges}. The flux distribution and the search strategy in this case then follows the one described in Section~\ref{sec:Search}.

For example, in the context of models where the \U1 generators follow the 4-1, 3-2 splitting,we find $\mathcal{O}(10^3)$ models which satisfy all anomaly conditions, in particular also \eqref{third}, and have all unwanted operators forbidden at tree level. Out of those, some are found to allow for suitable \U1 charges which lead to the desired operator structure. An example is the model given in Table~\ref{tab:BenchmarkBeyondRS}. Note that the analysis of the previous section still applies, since all \ten-curves have the same charge. 

In this benchmark model two singlets (or correspondingly appropriate instanton effects), denoted by $s_1$ and $s_2$, could generate the $\mu$-term and all Yukawa couplings. The charges for these fields have to be
\begin{align}
q(s_1)=(0,5)\,, \quad q(s_2)=(10,0)\, ,
\end{align}
which is similar to charges appearing in the benchmark models of the previous section. One slight difference is that one expects a higher suppression for the $\mu$-term since it is generated after two singlet insertions. Note that again we need a singlet with charge $10$ in the first \U1 to generate the desired coupling structure. Finally, we want to remark that we find significantly less phenomenologically appealing models when one of the additional U(1) symmetries results from a 5-0 split charge pattern (see Table \ref{tab:Splittings}). The reason for this might be that not only operators but the matter fields themselves have charges divisible by five, just as the singlet charges. This situation makes it difficult to retain a surviving symmetry which could help to control the dangerous couplings.

\section{Conclusions and outlook}
\label{sec:Conclusions}

In this paper we presented various interesting local F-theory models that have the potential to lead to a realistic GUT theory from string theory. For the first time we obtained models that satisfy all consistency conditions (four-dimensional anomaly cancellation and phenomenologically interesting couplings) and do not contain exotic matter.

We concentrated on F-theory \SU5 models which have up to two additional \U1 symmetries and allow for the three generations of quarks and leptons to arise from incomplete \SU5 representations. To find these appealing models we analyzed a class of models with toric sections whose geometric setup has been discussed recently in the literature, models arising within the spectral cover, and bottom-up models which feature \U1 charges of the same type, but lack an explicit geometric realization at this stage.

Our scan revealed phenomenologically interesting models in a subclass of the toric section constructions and in the bottom-up searches. A special feature of these models is that all matter from \ten-curves stay in complete multiplets which share the same charges under the additional \U1 symmetries, while in contrast the \fiveb-plets are usually split. This ``half-complete'' multiplet structure makes it possible to relate the charges of all operators among each other and is sufficient for generating phenomenologically interesting couplings. In particular, both types of models feature the top quark Yukawa coupling at tree level whereas the $\mu$-term and all baryon and lepton number violating operators are forbidden. We identified a singlet VEV configuration which promises to induce a realistic Yukawa structure, while all dimension four proton decay operators stay forbidden. We could further relate this situation to the presence of a residual matter parity.

At this stage we have postponed various phenomenologically relevant questions which have to be addressed when embedding the local models in a global string compactification. In particular we assumed a breakdown of the \SU5 gauge group to that of the Standard Model via hypercharge flux such that the hypercharge remains massless. In addition, we assumed that gauge coupling unification is not spoiled by large threshold corrections. For the subclass of models with toric sections and an explicit geometric realization we assumed an appropriate Green--Schwarz mechanism which can cancel the $\U1_Y$--$\U1_\alpha$--$\U1_\beta$ anomaly as motivated from the type IIB side. Note that this last anomaly is absent in our bottom-up models by construction and thus the last assumption need not be made there.

There are various interesting future research directions. It would be intriguing to find an explicit geometric realization of the bottom-up models presented here, and to construct compactifications that realize the above assumptions on hypercharge flux, gauge coupling unification and Green--Schwarz anomaly-cancellation. Given such a realization one of the next avenues might be to combine these local models with moduli stabilization. We hope to return to some of these questions in the near future.

\subsection*{Acknowledgments}
It is a pleasure to thank Hans Peter Nilles for collaboration at early stages of this project and for helpful comments and suggestions. Furthermore we want to thank Michael Blaszczyk, Herbi Dreiner, Rolf Kappl, Jan Keitel, Denis Klevers, Fernando Marchesano, Fernando Quevedo, and Timo Weigand for helpful discussions. This work was partially supported by the SFB-Transregio TR33 The Dark Universe (Deutsche Forschungsgemeinschaft) and the European Union 7th network program Unification in the LHC era (PITN-GA-2009-237920). The work of FR was supported by the German Science Foundation (DFG) within the Collaborative Research Center (SFB) 676 ``Particles, Strings and the Early Universe''.

\clearpage
\appendix

\section{Second benchmark model\label{sec:AppendixBenchmarkModel2}}
Here we present a second benchmark model based on the top $\tau_{5,2}$ whose spectrum is given in Table \ref{tab:TopCharges}. It has matter parity, but a slightly different Yukawa texture and dimension five proton decay operator structure, compared to the benchmark model from the main text.
\begin{table}[h!t]
\vspace{-0.2cm}
\centering
\renewcommand{\arraystretch}{1.4}
{\footnotesize
\begin{align*}
\begin{array}{|c|cc|cc|c|cc}
\cline{1-6}\cline{8-8}
\multicolumn{6}{l}{\text{\bf 1. Spectrum }} & & \multicolumn{1}{l}{\text{\bf 2. Singlet VEVs:}\,\, \boldsymbol{s_1\,, a}}\\
\cline{1-6}\cline{8-8}
\multicolumn{6}{l}{} & & \multirow{3}{*}{$q(s_1)=(0,5)\,,\quad q(a)=(10,0)\,.$}\\
\cline{1-6}
\text{Curve} & q_1 & q_2 & M & N & \text{Matter} & & \\ \cline{1-6}
\ten   & 1& 2& 3 & 0 & (Q+\bar{u}+\bar{e})_{1,2,3} & &\\
\cline{8-8}
\fiveb_1 & -3 & 4& 0 & 1& \bar{L}_{1}& & \multicolumn{1}{l}{\text{\bf 3. $\boldsymbol{\mu}$- and $\boldsymbol{\beta_i}$-terms}}\\
\cline{8-8}
\fiveb_3 & -3 & -1& 3 & -1& \bar{L}_{2,3}+\bar{d}_{1,2,3} & & \\
\fiveb_4 & 2 & 4 & 0 & -1 & H_u & & \multirow{3}{*}{\vspace{-3pt}$q(H_u L_i)=\begin{pmatrix}
 (-5,0) \\
 (-5,-5) \\
 (-5,-5)
\end{pmatrix} ,\ \quad q(H_u H_d) = (0,-5)\,.$}\\
\fiveb_5 & 2& -1& 0 & 1 & H_d & &\\
\cline{1-6}
\multicolumn{6}{l}{} & &  \\
\multicolumn{6}{l}{} & &  \\
\hline
\multicolumn{8}{l}{\text{\bf 4. Yukawa couplings}} \\
\hline
\multicolumn{8}{c}{} \\
\multicolumn{8}{c}{q(Q\bar{u} H_u) =(0,0)\, ,\quad\quad q(Q\bar{d} H_d) =
(0,0) \, ,\quad\quad q(\bar{e} L_j H_d)=\begin{pmatrix}
 (0,5) \\
 (0,0) \\
 (0,0)
\end{pmatrix}\,.} \\
\multicolumn{8}{c}{} \\
\hline
\multicolumn{8}{l}{\text{\bf 5. Allowed dimension five proton decay and Weinberg operators}} \\
\hline
\multicolumn{8}{c}{} \\
\multicolumn{8}{c}{q(\ten\, \ten\, \ten\, L_i) = \begin{pmatrix}
 (0,10) \\
 (0,5) \\
 (0,5)
\end{pmatrix} \, ,\quad  q(L_i\, L_j\, H_u\, H_u)= \begin{pmatrix}
(-10,0)&(-10,-5)&(-10,-5) \\
(-10,-5)& (-10,-10) & (-10,-10) \\
(-10,-5)& (-10,-10) & (-10,-10)
\end{pmatrix}\, ,} \\
\multicolumn{8}{c}{  q(\ten\, \ten\, \ten\, \bar{d})= (0,5)\,. } \\
\multicolumn{8}{c}{} \\
\hline
\multicolumn{8}{l}{\text{\bf 6. Forbidden operators}} \\
\hline
\multicolumn{8}{c}{} \\
\multicolumn{8}{c}{q(\bar{u}\bar{d}\bar{d})= (-5,0) \, ,\quad\quad q(\ten\, \ten\, d^*)= (5,5)\,.}
\end{array}
\end{align*}
}
\caption{Benchmark model 2 based on the top $\tau_{5,2}$ in detail, including the charges of all relevant operators.
}
\label{tab:BenchmarkB}
\end{table}

\clearpage
\begin{footnotesize}
\providecommand{\href}[2]{#2}\end{footnotesize}
\end{document}